\documentclass[10pt, a4paper, oneside, leqno]{article}
\usepackage{enumerate,amsthm,amsfonts, mathrsfs, amssymb, latexsym, dsfont}
\usepackage{setspace, fancyhdr}
\usepackage{amsmath}
\usepackage{avant}
\usepackage{url}
\usepackage[latin1]{inputenc}
\usepackage{verbatim}
\usepackage{graphicx}
\usepackage[left=3cm,right=3cm,top=4.5cm,bottom=3.5cm,headsep=2cm]{geometry}
\usepackage{color}
\usepackage{authblk}
\usepackage{microtype}

\newcommand{\Id}{\mathrm{I}}
\newcommand{\trace}{\mathrm{tr}}
\newcommand{\Dom}{\mathcal{O}}
\newcommand{\Val}{\mathcal{V}}
\newcommand{\ValRed}{\mathcal{U}}
\renewcommand{\L}{\mathcal{L}}

\newcommand{\prop}[1]{\underset{#1}{\sim}}
\newcommand{\ctrl}{[\underline\pi,\overline\pi]}
\newcommand{\adm}{\Pi}
\author[a]{Christoph~Belak}
\author[b]{An~Chen}
\author[c]{Carla~Mereu}
\author[d]{Robert~Stelzer}

\affil[a]{\small{TU Berlin, Institute of Mathematics, Stra\ss{}e des 17. Juni 136, 10623 Berlin, Germany. belak@math.tu-berlin.de}}
\affil[b]{\small{Ulm University, Institute of Insurance Science,  Helmholtzstrasse 20, 89081 Ulm, Germany. an.chen@uni-ulm.de}}
\affil[c]{\small{ccarlammereu@gmail.com}}
\affil[d]{\small{Ulm University, Institute of Mathematical Finance, Helmholtzstrasse 18, 89081 Ulm, Germany. robert.stelzer@uni-ulm.de}}

\title{Optimal investment with time-varying stochastic endowments}
\renewcommand{\tilde}{\widetilde}

\newcommand{\R}{\ensuremath{\mathbb{R}}}
\newcommand{\N}{\ensuremath{\mathbb{N}}}

\renewcommand{\P}{\ensuremath{\mathbb{P}}}
\newcommand{\F}{\ensuremath{\mathcal{F}}}

\newcommand{\ud}{\,\mathrm{d}}

\DeclareMathOperator{\E}{E}

\newcommand{\matlabcode}[3]%
{
\definecolor{number}{gray}{0.6}
\definecolor{keywords}{rgb}{1.0,0.3,0.3}
\definecolor{comments}{rgb}{0.1,0.65,0.1}
\definecolor{strings}{rgb}{0.3,0.0,1.0}
\lstset{language=Matlab,
        morekeywords={switch,case},
        sensitive=true,
        showspaces=false, 
        basicstyle=\ttfamily\small\mdseries,
      	keywordstyle=\bfseries\color{keywords},
 	      commentstyle=\color{comments},
 	      stringstyle=\color{strings},
 	      numbers=left,
 	      numberstyle=\scriptsize\color{number},
 	      stepnumber=1,
 	      breaklines=true,
 	      frame=none,
        showstringspaces=false,
 	      tabsize=4,
 	      xleftmargin=\bigskipamount,
 	      xrightmargin=\bigskipamount,
 	      aboveskip=\bigskipamount,
      	belowskip=0pt
}
\lstinputlisting[caption=#2, label=#3, frame=tb]{#1}
}

\def\ancomment#1{{{#1}}}

\newtheorem{Teo}{Theorem}[section]
\newtheorem{Lem}[Teo]{Lemma}
\newtheorem{Prop}[Teo]{Proposition}
\newtheorem{Cor}[Teo]{Corollary}
\theoremstyle{definition}

\theoremstyle{remark}
\newtheorem{oss}[Teo]{Remark}

\date{}
\allowdisplaybreaks[4]\numberwithin{equation}{section}
\begin{document}

\maketitle
\medskip
\hrule

\vspace{0.3cm}
{\small This paper considers a utility maximization and optimal asset allocation problem in the presence of a stochastic endowment that cannot be fully hedged through trading in the financial market. After studying continuity properties of the value function for general utility functions, we rely on the dynamic programming approach to solve the optimization problem for power utility investors including the empirically relevant and mathematically challenging case of relative risk aversion larger than one. For this, we argue that the value function is the unique viscosity solution of the Hamilton-Jacobi-Bellman (HJB) equation. The homogeneity of the value function is then used to reduce the HJB equation by one dimension, which allows us to prove that the value function is even a classical solution thereof. Using this, an optimal strategy is derived and its asymptotic behavior in the large wealth regime is discussed.

}
\vspace{0.3cm}
\smallskip
\hrule
\thispagestyle{empty}
\vspace{0.3cm}
In the present paper, we analyze the utility maximization or optimal investment problem of an economic agent who receives stochastic endowments over a finite time period. We deal with an incomplete market setting as the endowment risk is not perfectly correlated with the traded assets in the market. A specific example to motivate our model would be an economic agent who receives random salaries during her working life and invests a fixed proportion of the salaries in financial assets, optimizing the utility of her wealth at the retirement age. Another example could be a pension fund which receives random contributions from the members of the pension fund and invests to generate cash flows. In fact, the original idea of the paper arises from the study of defined contribution (DC) pension plans. This kind of pension scheme has become very popular over the last years and is substituting defined benefit pension plans (see e.g. \cite{broeders2010pension}).  \ancomment{In all OECD countries, defined contribution (DC) plans have played an increasingly important role in the last decades, see also the discussion in \cite{broeders2010pension} for further reasons for the transition from originally dominant defined benefit pension plans towards DC plans. In a DC plan, pension beneficiaries typically contribute a fraction of their salary to the fund.  Salaries or wages of the pension  beneficiaries, typically consisting of a fixed part and a performance-related  part (depending often on the company-wide, departmental as well as individual success), can be considered random. 
Likewise, it is reasonable to consider  contributions as being random  due to unforeseen changes in wages or salary, unemployment or reduced working hours, which has for example been faced by many workers due to the COVID pandemic.  However, a perfect correlation to the traded risky assets is obviously unrealistic. In other words, the salary/contribution risk is not fully hedgeable in the financial market.  In a DC plan, the pension beneficiaries bear the entire investment risk. For instance, DC beneficiaries in the U.S. manage the investment risk by so-called ``Individual Retirement Accounts'', or more frequently by making contributions to so-called 401(k) plans, see \cite{copeland2011target} and \cite{bosserhoff2021investment} for details on these plans. The stochastic optimization problem developed in this paper can be straightforwardly applied in this retirement saving problem.} 

The problem of maximizing the expected utility of an economic agent by investment and/or consumption dates back to \cite{merton1969} and \cite{Merton1971} and is further studied e.g.\ in \cite{cuoco1997optimal, duffie1997hedging, el1998optimization, koo1998consumption}, just to quote a few. Such problems have originally been solved using the dynamic programming approach which requires the assumption of Markovianity on the state process and leads to a Hamilton-Jacobi-Bellman (HJB) equation. In the literature, this approach is called primal approach. In the 1980s, researchers develop an alternative approach, the so called dual approach where the assumption of Markovian asset prices can be relaxed to solve the optimal investment problem. In a complete market setting the dual method has been studied e.g.\ in \cite{coxhuang1989, pliska1986} and in an incomplete market setting e.g.\ by \cite{hepearson1991a} and \cite{hepearson1991b}.

In the present paper, we consider the optimal asset allocation problem on a finite time horizon for an investor receiving an exogenous stochastic endowment modeled as a time-inhomogeneous geometric Brownian motion. We follow the primal HJB approach and argue that, in the case of power utility and under suitable regularity assumptions on the market parameters, the value function is a classical solution of the HJB equation. To obtain these results, we use a viscosity solution approach combined with classical existence results on regular solutions for nonlinear parabolic partial differential equations. More precisely, we first argue that the value function is the unique continuous viscosity solution of the HJB equation. Using the fact that the value function is homogeneous, we reduce the dimension of the HJB equation by one and obtain a fully nonlinear second-order parabolic equation which is uniformly elliptic in the spatial variable. From this, we are able to argue that the value function is even a classical solution of the HJB equation.

Proceeding from there, we then construct an optimal trading strategy. Since we restrict to strategies taking values in a compact-convex set, we shall see that it is a priori not clear if the stochastic differential equation (SDE) for the wealth process under the candidate optimal strategy (given in feedback form through the maximizer of the Hamiltonian in the HJB equation) admits a strong solution due to a lack of global Lipschitz properties of the drift and diffusion coefficients of the SDE. We are therefore forced to follow a non-classical approach to the construction of an optimal strategy: We first argue that an optimal strategy exists and then show that it can be represented in feedback form. In particular, this implies that the optimal strategy is unique. Given the optimal strategy, we finally analyze its asymptotic properties for large wealth-to-endowment ratios. We find that, asymptotically, the optimal strategy converges to the famous Merton ratio, which is the optimal strategy in the absence of random endowments.


The idea of incorporating a stochastic endowment in the classical utility maximization problem is indeed not new. Both dual and primal approaches (sometimes combined with Backward Stochastic Differential Equation (BSDE) techniques) are adopted to solve this optimization problem. Most of the papers in this field restrict themselves however to the analytically more tractable case of exponential utility functions (among several others \cite{davis2006optimal} and \cite{hu2005utility}). Some authors deal also with the problem of maximizing expected power utility of terminal wealth in the presence of exogenous endowments. In \cite{cvitanic2001utility}, the problem is solved via duality for a broad class of utility functions under the assumption that the random endowment is bounded. The existence and uniqueness of an optimal control are proven, but an explicit representation of the optimal strategy is subordinated to the decomposition of the elements of $(\mathrm{L}^\infty)^*$ into a regular and a singular part, which is hard to characterize explicitly. The authors of \cite{hugonnier2004optimal} overcome the problem and relax the hypothesis of boundedness by introducing the number of random endowments as a new control variable. \cite{Mostovyi2015} aims to examine under which conditions on the market model and on the utility stochastic field the maximizer of the optimal consumption problem exists. In addition, \cite{Mostovyi2015} looks into the properties of the value function and the corresponding dual problem.
\cite{MostovyiSirbu2020} studies existence and uniqueness of optimal investment and consumption strategies in a general incomplete semimartingale setting and provides a dual characterization in terms of an optional strong supermartingale deflator and a decreasing process. \cite{Zitkovic2011} focuses on examining the stability of the utility maximization problem with random endowment, i.e.\ when there is model and/or preference misspecification for the optimal investment problem. \cite{horst2014forward} extends the approach of \cite{hu2005utility} to the case of power utility functions, based on the martingale optimality principle combined with BSDE methods. They reduce the problem to the solution of a fully-coupled forward-backward stochastic differential equation, which is still not easy to solve. 

Most closely related to our paper are \cite{Bick2013} and \cite{duffie1997hedging}. The setting of \cite{Bick2013} is similar to ours and considers a utility maximization problem with stochastic income modelled by geometric Brownian motion. However, they find a near-optimal consumption and investment strategy which leads to a small wealth-equivalent loss compared to the unknown optimal strategy.  In \cite{duffie1997hedging}, the expected HARA utility from consumption is optimized over an infinite-time horizon and it yields an elliptic HJB equation, not depending on time. We adopt and extend their methods, developed in the elliptic case (infinite-time horizon), to the parabolic case (finite-time horizon). We further extend their model to a slightly broader class of random endowments, i.e., we allow for time-varying coefficients.
Most importantly, while \cite{duffie1997hedging} deals exclusively with the case of relative risk aversion less than one, we additionally extend the model to the empirically more relevant case of power utilities with relative risk aversion larger than one\footnote{
\ancomment{Note that there exist many empirical and experimental studies estimating the relative risk aversion (RRA) coefficients of individual investors. Although the resulting RRA levels differ among cases, all the levels seem to be larger than 1.   For example, \cite{barsky1997preference} finds that 65\% of their data shows a RRA above 3.76, 24\% below 2 and 12\% between 2 and 3.76. \cite{davies1981uncertain} describes 3, 4 and 5 as most reasonable estimates of RRA and \cite{azar2010bounds} estimates 3.01 and 3.74 as interval bounds for the RRA.}}. Especially the extension to the latter class of utility functions poses significant mathematical challenges in the viscosity characterization of the value function. In fact, we are able to establish a strong comparison principle for the (unreduced) HJB equation which covers the large relative risk aversion case as well, which is often a significant challenge in utility maximization problems; see, e.g., \cite{SonerVukelja2016,BelakEtAl2015}.

\ancomment{The key property which makes our model tractable enough to construct a classical solution of the HJB equation, and hence to construct an optimal trading strategy, is the homogeneity of the value function in the power utility case. Ultimately, as already observed in \cite{duffie1997hedging} in the infinite horizon optimal consumption problem, this property follows from the homogeneity of the utility function and the multiplicative dependence of the state processes on their initial values. We use the homogeneity of the value function to reduce the dimension of the state space by one, which is the key to establish the existence of a classical solution of the HJB equation. Note, however, that our reduced-form value function still depends on wealth through the wealth-to-endowment ratio, so even after the reduction our problem is fundamentally different to the setting in the classical work \cite{zariphopoulou} on optimizing expected utility in the presence of unhedgeable risk.}

\ancomment{Another advantage of our HJB approach is that it is highly amenable to numerical solutions and thus allows to investigate economic questions such as the qualitative behavior of the optimal strategy. We illustrate this by briefly comparing the optimal investment strategies of young and old (i.e.\ close to retirement) investors and by considering whether the widespread life cycle investment strategies following the principle ``the older you are, the less risky you should invest'' can be theoretically justified by our theory. }

The remainder of the paper is organized as follows. Section \ref{sec:model} describes the model setup and particularly the endowment process. In Section \ref{sec:stoch}, we introduce the optimization problem and study some of its basic properties such as well-posedness, continuity, concavity, and homogeneity. In Section~\ref{sec:hjb} we restrict ourselves to power utility and study properties of the associated HJB equation. Here, we show that the value \ancomment{function} is the unique viscosity solution of the HJB equation and prove that it is even a classical solution if the market parameters are sufficiently regular.  In Section \ref{sec:opt}, we construct an optimal strategy in the regular case and analyze its asymptotic behavior as the wealth-to-endowment ratio approaches infinity. \ancomment{In Section \ref{sec:num}, we apply numerical methods to illustrate concrete economical insights that can be gained.} Finally, Section \ref{sec:conclusion} concludes the paper.

\section{The model} \label{sec:model}
On a fixed filtered probability space $(\Omega, \F, \{\F_t\}_{t\in[0,T]}, \P)$ satisfying the usual hypotheses, consider a financial market consisting of a riskless and a risky asset. From now on let $T>0$ be a fixed finite time point.
Let $S^0$ and $S^1$ denote respectively the savings account and the risky asset. We assume that the two assets follow a Black-Scholes model:
\begin{equation*}
\ud S^0_t = rS^0_t\ud t,\qquad
\ud S^1_t = \mu S^1_t \ud t +\sigma S^1_t\ud W^1_t,
\end{equation*}
where $\mu,\ r\in\R,\ \sigma>0$ and $W^1$ is a Brownian motion on our filtered probability space.

The endowment process $c$ is assumed to have stochastic dynamics driven by another Brownian motion $W^C$ defined on the same probability space, which is assumed to be correlated with $W^1$ with a correlation coefficient\footnote{We exclude the perfect correlation cases $\rho= 1$ and $\rho=-1$. In these extreme cases, both the risky asset and the random income are fully driven by $W^1$. The optimization problem becomes simpler and differs from what we will present in the remaining text. For instance, in a retirement context, \cite{DybvigLiu2010} study the optimal consumption and retirement problem with stochastic labor income, where the labor income is perfectly correlated with the financial market risk.} $\rho\in(-1,1).$ In other words, we assume that there is another Brownian motion $W^2$ independent of $W^1$ such that $W^C = \rho W^1 + \sqrt{1-\rho^2}W^2$, which leads to the random endowment process $c$ given by
\begin{equation*}
\ud c_t = \mu_C(t)c_t\ud t + \sigma_C(t)c_t\ud W^C_t,\qquad t\in[0,T],\qquad c_0 = y,
\end{equation*}
where $y>0$ and $\mu_C : [0,T]\to \R$, $\sigma_C:[0,T]\to(0,\infty)$ are deterministic continuous functions.

\begin{oss}
Time-homogeneous geometric Brownian motions are sometimes chosen in the literature to model a stochastic income (see \cite{sundaresan1997valuation}). Thinking of the example of a DC pension scheme, we can interpret the process $c$ as a diffusion income, or rather a proportion of it possibly changing over time, which is paid continuously into the pension fund. From an analytical point of view, the fundamental feature of this family of random endowments is that, in the power utility case, the value function of the optimization problem turns out to be homogeneous in the spatial variables. This is the crucial property to achieve a reduction in the dimension of the problem.
\ancomment{Allowing for deterministic time varying drift and volatility does not substantially change the following mathematical analysis. It seems realistic that both drift and volatility of the income are somewhat higher at the beginning of a career than close to retirement. } 
\end{oss}

We assume that an agent with an initial wealth $x>0$ invests at any time $t$ a proportion $\pi_t$ of the wealth in the stock $S^1$ and $1-\pi_t$  in the risk-free asset $S^0$ with interest rate $r$. In addition, the random income is paid continuously to the account at rate $c_t$.  
The wealth process corresponding to the strategy $\pi$, denoted by $A^{\pi}$, is assumed to have the following dynamics:
\begin{equation*}
\ud A^\pi_t =\frac{\pi_tA^\pi_t}{S^1_t}\ud S^1_t + \frac{(1-\pi_t)A^\pi_t}{S^0_t}\ud S^0_t +c_t\ud t.
\end{equation*}
Defining $\theta := (\mu-r)/\sigma$, this can be written as
\begin{equation}\label{pension_income}
 \ud A^\pi_t =[(r + \sigma\theta\pi_t)A^\pi_t +c_t]\ud t + \sigma\pi_tA^\pi_t\ud W^1_t,\qquad t\in[0,T],\qquad A^\pi_0=x.
\end{equation}
This definition reflects the fact that the only additional cash injections are due to the continuous payments at rate $c_t$.

Note that in our set-up a strategy $\pi$ encodes the fraction of the overall wealth invested in the risky asset.
We assume that $\pi$ is a progressively measurable process taking values in a compact-convex set \ancomment{$[\underline\pi,\overline\pi]$} for real numbers $\underline\pi<\overline\pi$.
This includes the case of short-selling of the risky asset and borrowing of cash being restricted.
We write \ancomment{$\adm$} for the set of all such trading strategies.
To compare and discuss optimal trading in the presence and the absence of random endowments, we subsequently assume \ancomment{$\pi_M \in \ctrl$}, where $\pi_M$ denotes the Merton fraction
\begin{equation*}
 \pi_M :=\frac{\theta}{(1-\gamma)\sigma}.
\end{equation*}
It is a classical result that $\pi_M$ is the optimal strategy in the absence of random endowments, i.e.\ in case of $c\equiv 0$, and power utility with $1-\gamma$ being the relative risk aversion.
Let us highlight that our compactness assumption does not imply that the money invested in the risky asset or the cash position is bounded. Instead it means that there is a relative bound on the extent to which short-selling of the risky asset and taking credit on the bank account is allowed. As an example, choosing \ancomment{$\ctrl = [0,1]$} amounts to assuming that short-selling of the risky asset and borrowing of cash are prohibited.
Mathematically, the compactness assumption makes the study of the value function of the optimal investment problem significantly easier than in the unrestricted case. In fact, we shall make ample use of this assumption in our results in Section~\ref{sec:stoch} \ancomment{and in the viscosity characterization in Section~\ref{sec:hjb}; see also the discussion before Theorem~\ref{theoremComparison}}. In contrast, this trading constraint leads to significant challenges in the construction of an optimal strategy. In fact, even though we are able to show that the value function is a classical solution of the HJB equation and obtain a candidate optimal strategy in feedback form, it is unclear if the wealth process associated with the candidate optimal strategy exists; see the discussion in the beginning of Section~\ref{sec:opt} for details. We overcome this issue by first establishing an abstract existence result for optimal strategies and subsequently verifying that any optimal strategy coincides with our candidate optimal control.


\begin{oss}
Evidently, as \ancomment{$\ctrl$} is compact, any \ancomment{$\pi\in\adm$} is bounded. In particular, this ensures the existence and uniqueness of a solution for Equation \eqref{pension_income}.
\end{oss}

\begin{oss}In the definition of admissibility, one usually has to ensure that the wealth process never becomes negative. Under our assumptions, the wealth process stays positive without any
extra requirement on the admissible strategies. This can be seen from the fact that the explicit solution of~\eqref{pension_income} is given by
\begin{equation*}
 A^\pi_t = \left(x + \int_0^t \frac{c_s}{Z^\pi_s} \ud s\right)Z^\pi_t,\qquad t\in[0,T],
\end{equation*}
where $Z^\pi$ is a stochastic exponential factor given as
\begin{equation*}
 Z^\pi_t:= \exp\left\{\int_0^t\left[\pi_s\sigma\theta + r-\frac12(\pi_s\sigma)^2\right]\ud s + \int_0^t\sigma\pi_s\ud W^1_s\right\},\qquad t\in[0,T]. 
\end{equation*}
\end{oss}

\section{The optimization problem: statement and properties}\label{sec:stoch}

Recall that the Brownian motion $W^C$ represents the uncertainty in the income, which is supposed not to be  traded in the market. This makes the market incomplete and we are hence facing the problem of maximizing expected utility of an individual in an incomplete market. More precisely, we are looking for an optimal investment strategy \ancomment{$\pi^*\in\adm$} such that
\begin{equation*}
\E\bigl[U\bigl(A_T^{\pi^*} \bigr)\bigr]=\ancomment{\sup_{\pi\in\adm}}\E\bigl[U\bigl(A_T^\pi\bigr)\bigr],
\end{equation*}
where $U:(0,\infty)\to\R$ is a utility function which is twice continuously differentiable, increasing, concave, and  satisfies the Inada conditions $\lim_{x\uparrow+\infty}U'(x) = 0$ and  $\lim_{x\downarrow 0} U'(x) = \infty$.

Later on, we will restrict to the special case of $U$ being a constant relative risk aversion (CRRA) power utility function, i.e. for a risk aversion parameter $\gamma<1,$ $\gamma\neq 0$, we shall assume that $U = U_\gamma$ with
\begin{equation*}
 U_\gamma(x) = \frac{1}{\gamma}x^\gamma,\qquad x>0.
\end{equation*}
The power utility function is abundantly used in both theoretical and empirical research because of its analytical tractability. \ancomment{ Further, both empirical and experimental studies (e.g. \cite{barsky1997preference}) typically show that individuals demonstrate decreasing absolute risk aversion, and power utility falls into this category. In addition, the long-run behavior of the economy  suggests that the  risk aversion over long horizons, like for retirement decisions,  does not strongly depend on wealth, see \cite{campbell2002strategic}.  }

The agent is assumed to maximize expected utility of terminal wealth at the final time $T$, and thus the value function of the utility maximization problem is given by\footnote{A priori, it is of course not evident that $\Val$ is finite or even well-defined. We shall, however, see in Proposition~\ref{locbound} below that the positive part of $U(A^{\pi,t,x,y} _T)$ is integrable with a bound not depending on the trading strategy \ancomment{$\pi\in\adm$}, in which case $\Val$ is well-defined as an $\R\cup\{-\infty\}$-valued function.}
\begin{equation}\label{value_fct}
\Val:[0,T]\times\Dom \to\R\cup\{-\infty\},\qquad (t,x,y)\mapsto \Val(t,x,y) := \ancomment{\sup_{\pi\in\adm}}\E\left[U(A^{\pi,t,x,y} _T)\right],
\end{equation}
where $\Dom := (0,\infty)\times(0,\infty)$. In the definition of the value function, the state process is taken to be the pair $(A^{\pi,t,x,y},c^{t,y})$ and the notation $A^{\pi,t,x,y}$ and $c^{t,y}$ is used for the wealth and endowment process started at time $t$ in $x$ and $y$,	 respectively.

\subsection{A priori estimates, local boundedness, and continuity}

We first provide simple sufficient conditions under which the  value function is finite and continuous. A key ingredient is the following collection of a priori estimates on the state processes.

\begin{Lem}[A priori estimates]\label{LemmaAPriori}
Let $\alpha,\beta,\eta>0$. Then there exist $C_\alpha,C_\beta,C_\eta>0$ such that
\begin{align}
 \E\Bigl[\sup_{s\in[t,T]}\bigr|c_s^{t,y}\bigr|^\alpha\Bigr] &\leq C_\alpha |y|^\alpha, & (t,y)&\in[0,T]\times(0,\infty),\label{eqn_apriori_growth_endowment}\\
 \ancomment{\sup_{\pi\in\adm}}\E\Bigl[\sup_{s\in[t,T]}\bigr|A_s^{\pi,t,x,y}\bigr|^{-\beta}\Bigr] &\leq C_\beta |x|^{-\beta}, & (t,x,y)&\in[0,T]\times\Dom,\label{eqn_apriori_growth_negative}\\
 \ancomment{\sup_{\pi\in\adm}}\E\Bigl[\sup_{s\in[t,T]}\bigr|A_s^{\pi,t,x,y}\bigr|^\eta\Bigr] &\leq C_\eta |(x,y)|^\eta, & (t,x,y)&\in[0,T]\times\Dom.\label{eqn_apriori_growth_positive}
\end{align}
Moreover, there exists a constant $C>0$ such that
\begin{equation}\label{eqn_apriori_diff}
\ancomment{\sup_{\pi\in\adm}}\E\Bigl[ \bigl|A^{\pi,t,x,y}_T - A^{\pi,t',x',y'}_T\bigr|^2\Bigr] \leq C^2\Bigl(\bigl|(x,y)-(x',y')\bigr|^2 + |(x,y)|^2|t-t'|\Bigr)
\end{equation}
for all $(t,x,y),(t',x',y')\in[0,T]\times\Dom$.
\end{Lem}

\begin{proof}
The estimate~\eqref{eqn_apriori_diff} follows from a direct application of Proposition~3.22 in \cite{pardoux2014stochastic} (applying Eq.~(3.78)  with $p=2$, $q=\infty$, and $\lambda>1$ arbitrarily in the notation of \cite{pardoux2014stochastic}). To obtain~\eqref{eqn_apriori_growth_endowment}, we first observe that an application of It\^{o}'s lemma shows that the process $\tilde c := |c^{t,y}|^{\alpha/2}$ satisfies the (linear) Lipschitz SDE
\begin{equation*}
 \ud\tilde c_s = \tilde c_s\frac{\alpha}{2}\Bigl[\Bigl( \mu_C(s) + \frac14 (\alpha-2)\sigma_C(s)^2\Bigr)\ud s + \sigma_C(s)\ud W^C_s\Bigr],\qquad s\in[t,T],\qquad \tilde c_t = |y|^{\alpha/2}.
\end{equation*}
Classical a priori estimates for solutions of Lipschitz SDEs such as Corollary 2.10 in \cite{krylov2009controlled} applied to $\tilde c$ hence imply the existence of a constant $C_\alpha>0$ such that
\begin{equation*}
 \E\Bigl[\sup_{s\in[t,T]}\bigr|c_s^{t,y}\bigr|^\alpha\Bigr] = \E\Bigl[\sup_{s\in[t,T]}\bigr|\tilde c_s\bigr|^2\Bigr] \leq C_\alpha|\tilde c_t|^2 = C_\alpha|y|^\alpha.
\end{equation*}
Turning to~\eqref{eqn_apriori_growth_negative}, we use that $c^{t,y}\geq 0$ and $-\beta<0$ to estimate
\begin{equation}\label{apriori_proof_1}
\bigl|A_s^{\pi,t,x,y}\bigr|^{-\beta} = \Bigl( x + \int_t^s \frac{c_u^{t,y}}{Z^\pi_u} \ud u \Bigr)^{-\beta}\bigl|Z^\pi_s\bigr|^{-\beta} \leq |x|^{-\beta} \bigl|Z_s^\pi\bigr|^{-\beta} = |x|^{-\beta}\bigl|\tilde Z_s^\pi\bigr|^2,
\end{equation}
where, again by It\^{o}'s lemma, $\tilde Z^\pi := |Z^\pi|^{-\beta/2}$ solves the (linear) Lipschitz SDE
\begin{equation*}
\ud \tilde Z^\pi_s = -\frac{\beta}{2}\tilde Z^\pi_s\Bigl[\Bigl(r + \sigma\theta\pi_s - \frac{1}{4}(\beta+2) \sigma^2\pi_s^2\Bigr)\ud s + \sigma\pi_s\ud W^1_s\Bigr],\qquad s\in[t,T],\qquad \tilde Z_t = 1.
\end{equation*}
As before, we conclude that there exists a constant $C_\beta>0$ (independent of $\pi$ as the Lipschitz constant in the SDE for $\tilde Z^\pi$ can be chosen independent of $\pi$ since the set \ancomment{$\ctrl$} is bounded) such that
\begin{equation*}
\E\Bigl[\sup_{s\in[t,T]}\bigr|A_s^{\pi,t,x,y}\bigr|^{-\beta}\Bigr] \leq |x|^{-\beta}\E\Bigl[\sup_{s\in[t,T]} \bigl|\tilde Z_s^\pi\bigr|^2 \Bigr] \leq C_\beta |x|^{-\beta} |\tilde Z^\pi_t|^2 = C_\beta |x|^{-\beta}
\end{equation*}
and hence~\eqref{eqn_apriori_growth_negative} is obtained. Finally, regarding~\eqref{eqn_apriori_growth_positive}, we note that
\begin{equation*}
\E\Bigl[\sup_{s\in[t,T]}\bigr|A_s^{\pi,t,x,y}\bigr|^\eta\Bigr] \leq \E\Bigl[\sup_{s\in[t,T]}\bigr|\bigl(A_s^{\pi,t,x,y},c^{t,y}_s\bigr)\bigr|^\eta\Bigr] \leq \E\Bigl[\sup_{s\in[t,T]}\bigr|\bigl(A_s^{\pi,t,x,y},c^{t,y}_s\bigr)\bigr|^{2\vee\eta}\Bigr]^{(\eta/2)\wedge 1},
\end{equation*}
where we have to use Jensen's inequality in the last step if $\eta<2$. Applying again the a priori estimate for Lipschitz SDEs yields the existence of $\tilde C_\eta>0$ independet of $\pi$ such that
\begin{equation*}
\E\Bigl[\sup_{s\in[t,T]}\bigr|\bigl(A_s^{\pi,t,x,y},c^{t,y}_s\bigr)\bigr|^{2\vee\eta}\Bigr] \leq \tilde C_\eta |(x,y)|^{2\vee\eta},
\end{equation*}
which implies~\eqref{eqn_apriori_growth_positive} with $C_\eta := |\tilde C_\eta|^{(\eta/2)\wedge 1}$.
\end{proof}

The a priori estimates can be used to provide simple sufficient conditions under which the  value function is finite and $\E[U(A^{\pi,t,x,y} _T)]$ is bounded uniformly from below for all admissible strategies. As usual we define the positive and negative part of a function  $U^+(x):=\max\{U(x),0\}$ and $U^-(x):=\max\{-U(x),0\}$. Moreover, we remark that the Inada condition at infinity, i.e.\ $\lim_{x\uparrow\infty} U'(x) = 0$, implies that there exist constants $\eta\in(0,1]$ and $K_+ \geq 0$ such that
\begin{equation}\label{eqn_inada_estimate}
 U^+(x) \leq K_+\bigl(1 + |x|^\eta\bigr),\qquad x>0.
\end{equation}
E.g., in the case of power utility $U = U_\gamma$, this estimate holds for $(K_+,\eta) = (1/\gamma,\gamma)$ if $\gamma\in(0,1)$ and $K_+ = 0$ if $\gamma<0$.

\begin{Prop}[Local boundedness]\label{locbound}
\begin{enumerate}
\item Let $\eta\in(0,1]$ such that~\eqref{eqn_inada_estimate} holds. Then there exists a constant $C_+>0$ such that
\begin{equation*}
  \ancomment{\sup_{\pi\in\adm}} E\bigl[U^+\bigl(A_T^{\pi,t,x,y}\bigr)\bigr] \leq C_+\bigl(1 + |(x,y)|^\eta\bigr),\qquad (t,x,y)\in[0,T]\times\Dom.
\end{equation*}
\item Suppose that there exist constants $K_-\geq 0$ and $\beta>0$ such that $U^-(x) \leq K_-(1 + x^{-\beta})$. Then there exists a constant $C_->0$ such that
\begin{equation*}
\ancomment{\sup_{\pi\in\adm}} E\bigl[U^-\bigl(A_T^{\pi,t,x,y}\bigr)\bigr] \leq C_-\bigl(1 + |x|^{-\beta}\bigr),\qquad (t,x,y)\in[0,T]\times\Dom.
\end{equation*}
In particular, in this situation, the value function $\Val$ is locally bounded, i.e.\ there exists a constant $C_\Val>0$ such that
\begin{equation*}
 |\Val(t,x,y)| \leq C_\Val\bigl(1 + |x|^{-\beta} + |x|^\eta + |y|^\eta\bigr),\qquad (t,x,y)\in[0,T]\times\Dom.
\end{equation*}
\end{enumerate}
\end{Prop}

\begin{proof}
Let us fix $(t,x,y)\in\Dom$. To prove (i), we apply the estimate~\eqref{eqn_inada_estimate} followed by the a priori estimate~\eqref{eqn_apriori_growth_positive} to obtain the existence of a constant $C_\eta>0$ with
\begin{equation*}
\ancomment{\sup_{\pi\in\adm}}E\bigl[U^+(A_T^{\pi,t,x,y})\bigr] \leq K_+\Bigl(1  + \ancomment{\sup_{\pi\in\adm}}E\bigl[|A_T^{\pi,t,x,y}|^\eta\bigr]\Bigr) \leq K_+\Bigl( 1 + C_\eta|(x,y)|^\eta\Bigr).
\end{equation*}
The proof of (ii) works analogously by using~\eqref{eqn_apriori_growth_negative} in place of~\eqref{eqn_apriori_growth_positive}: There exists $C_\beta>0$ such that
\begin{equation*}
 \ancomment{\sup_{\pi\in\adm}} E\bigl[U^-\bigl(A_T^{\pi,t,x,y}\bigr)\bigr] \leq K_-\Bigl(1 + \ancomment{\sup_{\pi\in\adm}} E\Bigl[\bigl|A_T^{\pi,t,x,y}\bigr|^{-\beta}\Bigr]\Bigr)  \leq K_-\Bigl(1 + C_-|x|^{-\beta}\Bigr).
\end{equation*}
The estimate for $\Val$ is an immediate consequence of (i) and (ii).
\end{proof}

Note that for power utility functions $U_\gamma$ with $\gamma<0$ the assumption of part (ii) of the proposition is satisfied with $(K_-,\beta)=(-1/\gamma,-\gamma)$ and that for $\gamma<0$ the positive part and for $0 < \gamma < 1$ the negative part vanishes. The logarithmic utility $U = \log$ is an example of a common utility function for which the result applies and for which neither the positive nor negative part is zero.

In the next step, we derive continuity properties of the value function. This is done under the same assumptions as in the previous proposition together with a suitable growth condition on the derivative of the utility function.

\begin{Prop}[Continuity]\label{cont}
Suppose that there exist constants $K_-\geq 0$ and $\beta>0$ as well as constants $K'\geq 0$ and $\beta'>0$ such that 
\begin{equation}\label{eqnAssumpBdContinuous}
 U^-(x) \leq K_-\bigl(1 +  x^{-\beta}\bigr)\qquad\text{and}\qquad U'(x) \leq K'x^{-\beta'}\qquad\text{for all } x>0.
\end{equation}
Then we can find $C'>0$ such that
\begin{equation*}
|\Val(t,x,y)-\Val(\bar t,\bar x,\bar y)| \leq C'\min\{|x|,|\bar x|\}^{-\beta'}\Bigl(|(x,y)-(\bar x,\bar y)| + \max\bigl\{|(x,y)|,|(\bar x,\bar y)|\bigr\}|t-\bar t|^{1/2}\Bigr)
\end{equation*}
for all $(t,x,y),(\bar t,\bar x,\bar y)\in[0,T]\times\Dom$. In particular, $\Val$ is locally Lipschitz continuous in $(x,y)$ and locally $1/2$-H\"older continuous in $t$.
\end{Prop}

\begin{proof}
Let $(t,x,y),(\bar t,\bar x,\bar y)\in[0,T]\times\Dom$ and assume, without loss of generality, $\Val(t,x,y)\geq \Val(\bar t,\bar x,\bar y)$. For $\varepsilon>0$, we let \ancomment{$\pi\in\adm$} be an $\varepsilon$-optimal strategy in that
\begin{equation*}
\Val(t,x,y) \leq \E\bigl[U(A_T)\bigr] + \varepsilon,\qquad\text{where }A_T := A_T^{\pi,t,x,y}.
\end{equation*}
Note that $\pi$ exists since $\Val(t,x,y)>-\infty$. Moreover, Proposition~\ref{locbound} guarantees that
\begin{equation*}
\Val(\bar t,\bar x,\bar y) \geq \E\bigl[U(\bar A_T)\bigr] > -\infty,\qquad\text{where }\bar A_T := A_T^{\pi,\bar t,\bar x,\bar y}.
\end{equation*}
With this, using the concavity of $U$, the growth assumption on $U'$, and finally H\"{o}lder's inequality, we obtain the estimate
\begin{align}
|\Val(t,x,y) - \Val(\bar t,\bar x,\bar y)| = \Val(t,x,y) - \Val(\bar t,\bar x,\bar y)&\leq \E\bigl[U(A_T) - U(\bar A_T)\bigr] + \varepsilon\notag\\
&\leq \E\bigl[U'(\bar A_T)|A_T-\bar A_T|\bigr] + \varepsilon\notag\\
&\leq K'\E\bigl[|\bar A_T|^{-\beta'}|A_T-\bar A_T|\bigr] + \varepsilon\notag\\
&\leq K'\E\Bigl[|\bar A_T|^{-2\beta'}\Bigr]^{1/2}\E\Bigl[|A_T-\bar A_T|^2\Bigr]^{1/2} + \varepsilon.\label{eqn_proof_continuity_1}
\end{align}
Applying the a priori estimate~\eqref{eqn_apriori_growth_negative} now shows that
\begin{equation}\label{eqn_proof_continuity_2}
 \E\bigl[|\bar A_T|^{-2\beta'}\Bigr]^{1/2} \leq |C_{2\beta'}|^{1/2} |\bar x|^{-\beta'} \leq |C_{2\beta'}|^{1/2} \min\{|x|,|\bar x|\}^{-\beta'}
\end{equation}
and the a priori estimate~\eqref{eqn_apriori_diff} yields
\begin{align}
 \E\bigl[|A_T-\bar A_T|^2\bigr]^{1/2} &\leq C\Bigl(\bigl|(x,y)-(\bar x,\bar y)\bigr|^2 + |(x,y)|^2|t-\bar t|\Bigr)^{1/2}\notag\\
   &\leq C\Bigl(\bigl|(x,y)-(\bar x,\bar y)\bigr| + \max\bigl\{|(x,y)|,|(\bar x,\bar y)|\bigr\}|t-\bar t|^{1/2}\Bigr),\label{eqn_proof_continuity_3}
\end{align}
where we have used the subadditivity of the square root function for the last step. Note that neither of the constants $K',C_{2\beta'},C$ depends on the strategy $\pi$ and hence $C':= CK'|C_{2\beta'}|^{1/2}$ does not depend on $\pi$ either. Now plugging~\eqref{eqn_proof_continuity_2} and~\eqref{eqn_proof_continuity_3} into~\eqref{eqn_proof_continuity_1} leads to
\begin{equation*}
|\Val(t,x,y) - \Val(\bar t,\bar x,\bar y)| \leq C'\min\{|x|,|\bar x|\}^{-\beta'}\Bigl(\bigl|(x,y)-(\bar x,\bar y)\bigr| + \max\bigl\{|(x,y)|,|(\bar x,\bar y)|\bigr\}|t-\bar t|^{1/2}\Bigr) + \varepsilon.
\end{equation*}
The result follows as $\varepsilon$ is chosen arbitrarily and $C'$ does not depend on $\varepsilon$.
\end{proof}

Note that the power utility function satisfies the assumptions set forth in the previous proposition with $K' = 1$ and $\beta'=\gamma-1$. For later reference, we gather these results in a corollary.

\begin{Cor}[Power utility]
Suppose that $U$ is a power utility function, i.e.\ $U = U_\gamma$ for $\gamma<1$ with $\gamma\neq 0$. Then there exists a constant $C_\Val>0$ such that
\begin{equation}\label{eqnValPowerGrowth}
|\Val(t,x,y)| \leq C_\Val\Bigl(1 + |x|^{-\gamma^-} + |x|^{\gamma^+} + |y|^{\gamma^+}\Bigr),\qquad (t,x,y)\in[0,T]\times\Dom,
\end{equation}
where $\gamma^- := \max\{-\gamma,0\}$ and $\gamma^+ := \max\{\gamma,0\}$. Moreoever, $\Val$ is continuous and satisfies
\begin{equation*}
|\Val(t,x,y)-\Val(\bar t,\bar x,\bar y)| \leq C'\min\{|x|,|\bar x|\}^{1-\gamma}\Bigl(|(x,y)-(\bar x,\bar y)| + \max\bigl\{|(x,y)|,|(\bar x,\bar y)|\bigr\}|t-\bar t|^{1/2}\Bigr)
\end{equation*}
for all $(t,x,y),(\bar t,\bar x,\bar y)\in[0,T]\times\Dom$ for a constant $C'>0$.
\end{Cor}

\subsection{Further properties of the value function}

Before turning to the HJB equation associated with the optimization problem, we first study some further properties of the value function. In particular, we shall see that $\Val$ is concave and monotone in the spatial variables. Moreover, in the special case of power utility, we show that $\Val$ is homogeneous of degree $\gamma$.

\begin{Prop}[Monotonicity and concavity]\label{Prop_concavity}
$\Val$ is increasing and jointly concave in the spatial variables. In particular, for every fixed $t\in[0,T]$, the mapping $(x,y)\mapsto \Val(t,x,y)$ is locally Lipschitz continuous on the interior of its effective domain $\{(x,y)\in\Dom : \Val(t,x,y) > -\infty\}$. 
\end{Prop}

We note that, in Proposition~\ref{cont}, we have established a stronger continuity result. However, for this we have to assume~\eqref{eqnAssumpBdContinuous}, whereas Proposition~\ref{Prop_concavity} is valid without additional assumptions.

\begin{proof}[Proof of Prop.~\ref{Prop_concavity}]
The proof  works along standard lines (cf., e.g., Section 3.6.1 in \cite{pham2009continuous}). Note that continuity follows immediately from concavity, as every concave function is locally Lipschitz continuous in the interior of its effective domain.
	
Step 1: Monotonicity. Let us fix $t\in[0,T]$, $(x_1,y_1),(x_2,y_2)\in\Dom$ with $x_1\leq x_2$ and $y_1\leq y_2$, and a strategy \ancomment{$\pi\in\adm$}. Since the endowment process is a geometric Brownian motion we see that
\begin{equation*}
c_T^{t,y_1} = y_1 c_T^{t,1} \leq  y_2 c_T^{t,1} = c_T^{t,y_2}
\end{equation*}
and from the explicit representation of the wealth process we find that
\begin{equation*}
 A_T^{\pi,t,x_1,y_1} = \Bigl(x_1 + \int_t^T \frac{c_s^{t,y_1}}{Z_s^\pi} \ud s\Bigr)Z_T^\pi \leq \Bigl(x_2 + \int_t^T \frac{c_s^{t,y_2}}{Z_s^\pi} \ud s\Bigr)Z_T^\pi =  A_T^{\pi,t,x_2,y_2}.
\end{equation*}
Since $\pi$ was chosen arbitrarily and the utility function $U$ is monotone, this shows that
\begin{equation*}
 \Val(t,x_1,y_1) = \ancomment{\sup_{\pi\in\adm}} \E\bigl[U\bigl(A_T^{\pi,t,x_1,y_1}\bigr)\bigr] \leq \ancomment{\sup_{\pi\in\adm}} \E\bigl[U\bigl(A_T^{\pi,t,x_2,y_2}\bigr)\bigr] = \Val(t,x_2,y_2),
\end{equation*}
i.e.\ $\Val$ is monotone in both its second and third argument.

Step 2: Concavity. Let us again fix $t\in[0,T]$ and $(x_1,y_1),(x_2,y_2)\in\Dom$. Let moreover \ancomment{$\pi^1,\pi^2\in\adm$}, choose $\vartheta\in[0,1]$, and define
\begin{equation*}
 x_\vartheta := \vartheta x_1 + (1-\vartheta)x_2\qquad\text{and}\qquad y_\vartheta := \vartheta y_1 + (1-\vartheta) y_2.
\end{equation*}
Moreover, with $A^i := A^{\pi^i,t,x_i,y_i}$, $i=1,2$, we define
\begin{equation*}
 A^\vartheta := \vartheta A^1 + (1-\vartheta)A^2\qquad\text{and}\qquad \pi^\vartheta := \frac{\vartheta A^1\pi^1 + (1-\vartheta)A^2\pi^2}{A^\vartheta}.
\end{equation*}
As $\pi^\vartheta$ is a convex combination of $\pi^1$ and $\pi^2$, it follows that $\pi^\vartheta$ is \ancomment{$\ctrl$-valued} and hence $\pi^\vartheta$ is admissible. Let us \ancomment{show} that $A^\vartheta$ is the wealth process corresponding to $\pi^\vartheta$. For this, observe that
\begin{equation*}
 c^{t,y_\vartheta} = y_\vartheta c^{t,1} = \vartheta y_1c^{t,1} + (1-\vartheta)y_2 c^{t,1} = \vartheta c^{t,y_1} + (1-\vartheta)c^{t,y_2},
\end{equation*}
implying that
\begin{align*}
\ud A^\vartheta_s &= \vartheta \ud A^1_s + (1-\vartheta)\ud A^2_s\\
    &= r\bigl[\vartheta A^1_s + (1-\vartheta)A^2_s\bigr]\ud s + \sigma\theta\bigl[\vartheta \pi^1_sA^1_s + (1-\vartheta)\pi^2_sA^2_s \bigr]\ud s + \bigl[\vartheta c_s^{t,y_1} + (1-\vartheta)c_s^{t,y_2}\bigr]\ud s\\
    &\hspace{8.5cm} + \sigma\bigl[\vartheta \pi^1_sA^1_s + (1-\vartheta)\pi^2_sA^2_s \bigr]\ud W^1_s\\
    &=\bigl[ \bigl(r + \sigma\theta \pi^\vartheta_s\bigr) A^\vartheta_s + c^{t,y_\vartheta}_s\bigr]\ud s + \sigma \pi^\vartheta_sA^\vartheta_s\ud W^1_s
\end{align*}
and hence uniqueness of solutions of linear SDEs implies that $A^{\pi^\vartheta,t,x_\vartheta,y_\vartheta} = A^\vartheta = \vartheta A^1 + (1-\vartheta)A^2$, i.e.\ $A^\vartheta$ is the wealth process corresponding to $\pi^\vartheta$ started in $(t,x_\vartheta,y_\vartheta)$.  Using the concavity of the utility function, we get
\begin{equation*}
\Val(t,x_\vartheta,y_\vartheta) \geq \E\bigl[U(A^\vartheta_T)\bigr] \geq \vartheta \E\bigl[U(A^1_T)\bigr] + (1-\vartheta) \E\bigl[U(A^2_T)\bigr],
\end{equation*}
but since $\pi^1,\pi^2$ are chosen arbitrary, taking the supremum on the right hand side, we arrive at
\begin{equation*}
\Val(t,x_\vartheta,y_\vartheta)\geq\vartheta \Val(t,x_1,y_1)+(1-\vartheta)\Val(t,x_2,y_2).\qedhere
\end{equation*}
\end{proof}

We conclude this section by showing that, in the power utility case $U = U_\gamma$, the value function is homogeneous of degree $\gamma$ in the spatial variables.

\begin{Lem}[Homogeneity]\label{gen_homog_red}
Suppose that $U=U_\gamma$ with $\gamma<1$, $\gamma\neq 0$. Then the value function $\Val$ is homogeneous of degree $\gamma$ in the spatial variables, i.e.\
\begin{equation*}
\Val(t,kx,ky) = k^\gamma \Val(t,x,y)\qquad\text{for all } k>0\text{ and } (t,x,y)\in[0,T]\times\Dom.
\end{equation*}
As a consequence, $\Val$ can be represented in separable form as
\begin{equation*}
 \Val(t,x,y) = y^\gamma \ValRed\bigl(t, x/y\bigr),\qquad (t,x,y)\in[0,T]\times\Dom,
\end{equation*}
where
\begin{equation}\label{eqnValRed}
 \ValRed:[0,T]\times(0,\infty)\to\R,\qquad (t,z)\mapsto \ValRed(t,z) := \Val(t,z,1).
\end{equation}
\end{Lem}

\begin{proof}
The separable representation of $\Val$ follows immediately from homogeneity with $k=y$ since
\begin{equation*}
 \Val(t,x,y) = \Val\bigl(t,y(x/y),y\bigr) = y^\gamma \Val\bigl(t,x/y,1\bigr),\qquad (t,x,y)\in[0,T]\times\Dom.
\end{equation*}
To see that $\Val$ is homogeneous, fix $(t,x,y)\in[0,T]\times\Dom$, $k>0$, and \ancomment{$\pi\in\adm$}. Since $c^{t,ky} = k c^{t,y}$ we find that
\begin{equation*}
 A^{\pi,t,kx,ky}_T = \Bigl(kx + \int_t^T \frac{c_s^{t,ky}}{Z^\pi_s}\ud s\Bigr)Z^\pi_T = k\Bigl(x + \int_t^T \frac{c_s^{t,y}}{Z^\pi_s}\ud s\Bigr)Z^\pi_T = kA^{\pi,t,x,y}_T. 
\end{equation*}
But then $\Val$ must be homogeneous of degree $\gamma$ since
\begin{equation*}
\Val(t,kx,ky) = \ancomment{\sup_{\pi\in\adm}}\E\bigl[U_\gamma\bigl(A_T^{\pi,t,kx,ky}\bigr)\bigr] = \ancomment{\sup_{\pi\in\adm}}\E\Bigl[\frac{1}{\gamma}\bigl(k A_T^{\pi,t,x,y}\bigr)^\gamma\Bigr] = k^\gamma \Val(t,x,y).\qedhere
\end{equation*}
\end{proof}

\section{The Hamilton-Jacobi-Bellman equation}\label{sec:hjb}

Henceforth, we restrict our attention to power utility, i.e.\ $U = U_\gamma$. Applying well-known principles of stochastic control, see e.g.\ Chapter 3 in \cite{pham2009continuous}, we can write down the HJB equation for the value function of our control problem:
\begin{align}
 - \Val_t(t,x,y) - \ancomment{\sup_{\pi\in\ctrl}}\bigl\{\L^\pi \Val(t,x,y)\bigr\} &= 0, & (t,x,y) &\in[0,T)\times\Dom,\label{HJB_cont}\\
 \Val(t,x,y) &= U_\gamma(x), & (x,y) &\in\Dom,\notag
\end{align}
where the linear differential operator $\L^\pi$ is defined through
\begin{equation*}
\L^\pi \Val(t,x,y) := \bigl[(r + \sigma\theta\pi)x + y\bigr]\Val_x + \mu_Cy\Val_y  + \frac12\sigma^2\pi^2x^2\Val_{xx} + \frac12\sigma_C^2y^2\Val_{yy} + \rho\sigma\sigma_C\pi xy\Val_{xy}.
\end{equation*}
Here, and in the following, we denote partial derivatives by subscripts and often omit the argument of functions in the equations. The aim of this section is to characterize the value function as the unique viscosity solution of the HJB equation. Moreover, we show that $\Val$ is even a classical solution provided that $\mu_C$ and $\sigma_C$ are continuously differentiable.

\subsection{A family of supersolutions}

Before linking the value function to the HJB equation, let us first study the existence of classical supersolutions of the HJB equation. These play a crucial role for our subsequent analysis and serve, in particular, as abstract boundary/growth conditions for the HJB equation.

We begin by defining two constants
\begin{equation*}
 K:= r + \frac{\theta^2}{2(1-\gamma)}
\end{equation*}
as well as
\begin{equation*}
\lambda :=\max\Bigl\{ -  r - \ancomment{\inf_{\pi\in\ctrl}}\bigl\{\sigma\theta\pi  - \frac{1}{2}(1+\kappa)\sigma^2\pi^2 \bigr\} , -\inf_{t\in[0,T]}\bigl\{\mu_C(t)- \frac{1}{2}(1+\kappa)\sigma_C(t)^2\bigr\}\Bigr\}.
\end{equation*}
With this, we introduce a parametric family of functions
\begin{equation*}
\psi_{\gamma,\varepsilon,\iota} : [0,T]\times\Dom\to\R,\qquad (t,x,y)\mapsto\psi_{\gamma,\varepsilon,\iota}(t,x,y)
\end{equation*}
for $\gamma<1$ with $\gamma\neq 0$, $\varepsilon\in[0,1]$, and $\iota\in\{0,1\}$. We assume that $\psi_{\gamma,\varepsilon,\iota}$ takes the form
\begin{equation*}
 \psi_{\gamma,\varepsilon,\iota}(t,x,y) := U_\gamma\Bigl(x + y\varphi_\iota(t) + \varepsilon e^{-r(T-t)} \Bigr) e^{\gamma K(T-t)} + \iota\bigl(x^{-(1+|\gamma|)} + y^{-(1+|\gamma|)}\bigr)e^{(1+|\gamma|)\lambda(T-t)}
\end{equation*}
for all $(t,x,y)\in[0,T]\times\Dom$, where the function $\varphi_\iota:[0,T]\to\R$ is given as the unique solution of the ordinary differential equation
\begin{equation*}
\dot \varphi_\iota(t) = \bigl(r-\mu_C(t)+\rho\theta\sigma_C(t)\bigr)\varphi_\iota(t) - 1,\qquad t\in[0,T],\qquad \varphi_\iota(T) = \iota.
\end{equation*}
Observe that $\varphi_\iota(t) > 0$ unless $\iota=0$ and $t=T$. In any case, $\varphi_\iota$ is nonnegative and hence, in particular, $\psi_{\gamma,\varepsilon,\iota}$ well-defined. We proceed to show that $\psi_{\gamma,\varepsilon,\iota}$ is a classical supersolution of the HJB equation.

\begin{Prop}[Classical supersolution]
For any choice of $\gamma<1$ with $\gamma\neq 0$, $\varepsilon\in[0,1]$, and $\iota\in\{0,1\}$, the function $\psi:=\psi_{\gamma,\varepsilon,\iota}$ is a supersolution of the HJB equation, i.e.
\begin{equation*}
-\psi_t(t,x,y) - \ancomment{\sup_{\pi\in\ctrl}}\bigl\{\L^\pi\psi(t,x,y)\bigr\} \geq 0,\qquad (t,x,y)\in[0,T)\times\Dom.
\end{equation*}
Moreover, it is a strict supersolution if $\iota = 1$.
\end{Prop}

\begin{proof}
We decompose $\psi$ as $\psi = \psi^1 + \psi^2$, where
\begin{align*}
\psi^1(t,x,y) &:= U_\gamma\Bigl(x + y\varphi_\iota(t) + \varepsilon e^{-r(T-t)} \Bigr) e^{\gamma K(T-t)}, && (t,x,y)\in[0,T]\times\Dom,\\
\psi^2(t,x,y) &:= \iota\bigl(x^{-(1+|\gamma|)} + y^{-(1+|\gamma|)}\bigr)e^{(1+|\gamma|)\lambda(T-t)}, && (t,x,y)\in[0,T]\times\Dom.
\end{align*}
By linearity of the operator $\L^\pi$, it follows that
\begin{equation*}
-\psi_t - \ancomment{\sup_{\pi\in\ctrl}} \bigl\{ \L^\pi \psi\bigr\} \geq \Bigl(-\psi^1_t - \ancomment{\sup_{\pi\in\ctrl}} \bigl\{ \L^\pi \psi^1\bigr\}\Bigr) + \Bigl( -\psi^2_t - \ancomment{\sup_{\pi\in\ctrl}} \bigl\{ \L^\pi \psi^2\bigr\}\Bigr).
\end{equation*}
Hence, in order to conclude, it suffices to show that $\psi^1$ and $\psi^2$ are supersolutions, and $\psi^2$ is a strict supersolution if $\iota = 1$.

Step 1: $\psi^2$ is a (strict) supersolution. If $\iota = 0$, then $\psi^2\equiv 0$ and it is thus a solution of the HJB equation. Let us hence assume that $\iota=1$ and prove that, in this case, $\psi^2$ is a strict supersolution. Setting $\kappa:=(1+|\gamma|)$, the partial derivatives of $\psi^2$ are given by
\begin{align*}
\psi^2_t &= -\kappa\lambda\bigl(x^{-\kappa}+y^{-\kappa}\bigr)e^{\kappa\lambda(T-t)}, &
\psi^2_x &= -\kappa x^{-\kappa-1} e^{\kappa\lambda(T-t)}, &
\psi^2_y &= -\kappa y^{-\kappa-1} e^{\kappa\lambda(T-t)},\\
\psi^2_{xx} &= \kappa(1+\kappa) x^{-\kappa-2} e^{\kappa\lambda(T-t)}, &
\psi^2_{yy} &= \kappa(1+\kappa) y^{-\kappa-2} e^{\kappa\lambda(T-t)}, &
\psi^2_{xy} &= 0.
\end{align*}
Plugging these derivatives into the HJB equation and rearranging terms yields
\begin{align*}
-\psi^2_t - \ancomment{\sup_{\pi\in\ctrl}}\bigl\{\L^\pi\psi^2\bigr\}
&= \kappa\ancomment{\inf_{\pi\in\ctrl}}\Bigl\{\lambda  + r + \sigma\theta\pi - \frac{1}{2}(1+\kappa)\sigma^2\pi^2 \Bigr\}x^{-\kappa} e^{\kappa\lambda(T-t)} + \kappa yx^{-\kappa-1}e^{\kappa\lambda(T-t)}\\
 &\hspace{4.5cm}+\kappa \Bigl(\lambda  + \mu_C(t) - \frac{1}{2}(1+\kappa)\sigma_C(t)^2\Bigr) y^{-\kappa}e^{\kappa\lambda(T-t)}\\
 &\geq \kappa\Bigl(\lambda  +  r + \ancomment{\inf_{\pi\in\ctrl}}\Bigl\{\sigma\theta\pi  - \frac{1}{2}(1+\kappa)\sigma^2\pi^2 \Bigr\}\Bigr)x^{-\kappa}e^{\kappa\lambda(T-t)}  + \kappa yx^{-\kappa-1}e^{\kappa\lambda(T-t)}\\
  &\hspace{3cm} + \kappa\Bigl(\lambda  + \inf_{t\in[0,T]}\Bigl\{\mu_C(t)- \frac{1}{2}(1+\kappa)\sigma_C(t)^2\Bigr\}\Bigr) y^{-\kappa}e^{\kappa\lambda(T-t)}.
\end{align*}
By the choice of $\lambda$, the terms inside the brackets are nonnegative and hence
\begin{equation*}
-\psi^2_t - \ancomment{\sup_{\pi\in\ctrl}}\bigl\{\L^\pi\psi^2\bigr\} \geq \kappa yx^{-\kappa-1} e^{\kappa\lambda(T-t)} > 0,
\end{equation*}
i.e.\ $\psi^2$ is a strict supersolution of the HJB equation.

Step 2: $\psi^1$ is a supersolution. Writing $\varphi = \varphi_\iota$ and using the ordinary differential equation for $\varphi$, the partial derivatives of $\psi^1$ are given by
\begin{align*}
\psi^1_t &= - K\Bigl(x + y\varphi(t) + \varepsilon e^{-r(T-t)} \Bigr)^\gamma e^{\gamma K(T-t)}\\
                        &\hspace{0.5cm} + \Bigl(\bigl(r-\mu_C(t)+\rho\theta\sigma_C(t)\bigr)y\varphi(t) - y + r\varepsilon e^{-r(T-t)}\Bigr)\Bigl(x + y\varphi(t) + \varepsilon e^{-r(T-t)} \Bigr)^{\gamma-1} e^{\gamma K(T-t)},\\
\psi^1_x &= \Bigl(x + y\varphi(t) + \varepsilon e^{-r(T-t)} \Bigr)^{\gamma-1 } e^{\gamma K(T-t)},\\
\psi^1_y &= \varphi(t)\Bigl(x + y\varphi(t) + \varepsilon e^{-K(T-t)} \Bigr)^{\gamma-1} e^{\gamma K(T-t)},\\
\psi^1_{xx} &= -(1-\gamma)\Bigl(x + y\varphi(t) + \varepsilon e^{-r(T-t)} \Bigr)^{\gamma-2 } e^{\gamma K(T-t)},\\
\psi^1_{yy} &= -(1-\gamma)\varphi(t)^2\Bigl(x + y\varphi(t) + \varepsilon e^{-r(T-t)} \Bigr)^{\gamma-2} e^{\gamma K(T-t)},\\
\psi^1_{xy} &= -(1-\gamma)\varphi(t)\Bigl(x + y\varphi(t) + \varepsilon e^{-r(T-t)} \Bigr)^{\gamma-2 } e^{\gamma K(T-t)}.
\end{align*}
Plugging these derivatives into the HJB equation and rearranging terms yields
\begin{align*}
&\mathrel{\phantom{=}} -\psi^1_t(t,x,y) - \ancomment{\sup_{\pi\in\ctrl}}\bigl\{\L^\pi\psi^1(t,x,y)\bigr\}\\
&= - e^{\gamma K(T-t)}\ancomment{\sup_{\pi\in\ctrl}}\Bigl\{- K\Bigl(x + y\varphi(t) + \varepsilon e^{-r(T-t)} \Bigr)^\gamma\\
&\hspace{1cm} + \Bigl[\bigl(r + \sigma\theta\pi\bigr) x + \bigl(r+\rho\theta\sigma_C(t)\bigr)y\varphi(t)  + r\varepsilon e^{-r(T-t)}\Bigr]\Bigl(x + y\varphi(t) + \varepsilon e^{-r(T-t)} \Bigr)^{\gamma-1} \\
&\hspace{1cm} - \frac12(1-\gamma)\Bigl[\sigma^2\pi^2x^2 + 2\rho\sigma\sigma_C(t)\pi xy\varphi(t) + \sigma_C(t)^2y^2\varphi(t)^2\Bigr]\Bigl(x + y\varphi(t) + \varepsilon e^{-r(T-t)} \Bigr)^{\gamma-2 }\Bigr\}.
\end{align*}
Instead of maximizing over \ancomment{$\ctrl$} we can maximize over all of $\R$, making the last expression smaller in doing so. The maximizer in the supremum is given by
\begin{equation*}
 \hat\pi := \frac{1}{(1-\gamma)\sigma x}\Bigl[\theta \Bigl(x + y\varphi(t) + \varepsilon e^{-r(T-t)} \Bigr)  - (1-\gamma)\rho\sigma_C(t) y\varphi(t)\Bigr].
\end{equation*}
Plugging this into the above equation and rearranging terms again then yields
\begin{align*}
&\mathrel{\phantom{=}} -\psi^1_t(t,x,y) - \ancomment{\sup_{\pi\in\ctrl}}\bigl\{\L^\pi\psi^1(t,x,y)\bigr\}\\
&\geq - e^{\gamma K(T-t)}\Bigl[- K\Bigl(x + y\varphi(t) + \varepsilon e^{-r(T-t)} \Bigr)^\gamma\\
&\hspace{1cm} + \Bigl[\bigl(r + \sigma\theta\hat\pi\bigr) x + \bigl(r+\rho\theta\sigma_C(t)\bigr)y\varphi(t)  + r\varepsilon e^{-r(T-t)}\Bigr]\Bigl(x + y\varphi(t) + \varepsilon e^{-r(T-t)} \Bigr)^{\gamma-1} \\
&\hspace{1cm} - \frac12(1-\gamma)\Bigl[\sigma^2\hat\pi^2x^2 + 2\rho\sigma\sigma_C(t)\hat\pi xy\varphi(t) + \sigma_C(t)^2y^2\varphi(t)^2\Bigr]\Bigl(x + y\varphi(t) + \varepsilon e^{-r(T-t)} \Bigr)^{\gamma-2 }\Bigr]\\
&= \Bigl(K - r - \frac{\theta^2}{2(1-\gamma)}\Bigr)\Bigl(x + y\varphi(t) + \varepsilon e^{-r(T-t)} \Bigr)^\gamma e^{\gamma K(T-t)}\\
&\hspace{3cm} + \frac12(1-\gamma)(1-\rho)^2\sigma_C(t)^2 y^2\varphi(t)^2\Bigl(x + y\varphi(t) + \varepsilon e^{-r(T-t)} \Bigr)^{\gamma-2 } e^{\gamma K(T-t)} \geq 0,
\end{align*}
where the nonnegativity follows from the choice of $K$.
\end{proof}

With the existence of classical supersolutions, we can derive tight bounds on the value function.

\begin{Prop}[Tight bounds on $\Val$]\label{propTightBounds}
The value function $\Val$ satisfies
\begin{equation*}
U_\gamma(x)e^{\gamma K (T-t)}\leq \Val(t,x,y) \leq U_\gamma\bigl(x + y\varphi_0(t)\bigr)e^{\gamma K(T-t)},\qquad (t,x,y)\in[0,T]\times\Dom.
\end{equation*}
\end{Prop}

\begin{proof}
Step 1: We first prove the lower bound. For this, consider the constant strategy \ancomment{$\pi \equiv \pi_M\in\adm$} and recall that $A_T^{\pi_M,t,x,y} \geq A_T^{\pi_M,t,x,0} = x A_T^{\pi_M,t,1,0}$. But then
\begin{equation*}
\Val(t,x,y) \geq \E\bigl[U_\gamma\bigl(A_T^{\pi_M,t,x,y}\bigr)\bigr] \geq \E\bigl[U_\gamma\bigl(xA_T^{\pi_M,t,1,0}\bigr)\bigr] = U_\gamma(x)\E\bigl[\bigl(A_T^{\pi_M,t,1,0}\bigr)^\gamma\bigr].
\end{equation*}
The last expectation can be computed explicitly and is given by
\begin{multline*}
\E\bigl[\bigl(A_T^{\pi_M,t,1,0}\bigr)^\gamma\bigr] = \E\Bigl[e^{ \gamma(r+\sigma\theta\pi_M - \frac12\sigma^2\pi_M^2) (T-t) + \gamma\sigma\pi_M (W^1_T-W^1_t)}\Bigr]\\
 = e^{ \gamma(r+\sigma\theta\pi_M - \frac12(1-\gamma)\sigma^2\pi_M^2) (T-t)} = e^{\gamma K(T-t)}.
\end{multline*}

Step 2: To prove the upper bound, let $(t,x,y)\in[0,T]\times\Dom$ as well as \ancomment{$\pi\in\adm$} and write $A := A^{\pi,t,x,y}$ and $c := c^{t,y}$. For $\varepsilon\in(0,1]$ fixed, set $\psi^\varepsilon := \psi_{\gamma,\varepsilon,0}$ and denote by $\rho_n$, $n\in\N$, a localizing sequence of the local martingale
\begin{equation*}
 M_s := \int_t^s \sigma\pi_u A_u \psi^\varepsilon_x(u,A_u,c_u)\ud W^1_u + \int_t^s \sigma_C(u)c_u\psi^\varepsilon_y(u,A_u,c_u)\ud W^C_u,\qquad s\in[t,T].
\end{equation*}
Applying It\^{o}'s lemma to $\psi^\varepsilon(\rho_n,A_{\rho_n},c_{\rho_n})$ and using the fact that $\psi^\varepsilon$ is a supersolution of the HJB equation, it follows that
\begin{align*}
 \psi^\varepsilon(t,x,y) &= \E\Bigl[\psi^\varepsilon(\rho_n,A_{\rho_n},c_{\rho_n}) + \int_t^{\rho_n}\bigl[ - \psi^\varepsilon_t(s,A_s,c_s) -  \L^{\pi_s}\psi^\varepsilon(s,A_s,c_s)\bigr]\ud s - M_{\rho_n}\Bigr]\\
   &\geq \E\bigl[\psi^\varepsilon(\rho_n,A_{\rho_n},c_{\rho_n})\bigr].
\end{align*}
As $\varepsilon>0$, we see that $\psi^\varepsilon$ is lower bounded. Hence we may apply Fatou's lemma to arrive at
\begin{equation*}
 \psi^\varepsilon(t,x,y) \geq \liminf_{n\to\infty} \E\bigl[\psi^\varepsilon(\rho_n,A_{\rho_n},c_{\rho_n})\bigr] \geq \E\bigl[\psi^\varepsilon(T,A_T,c_T)\bigr] = \E\bigl[U_\gamma\bigl(A_T + \varepsilon\bigr)\bigr] \geq \E\bigl[U_\gamma\bigl(A_T\bigr)\bigr].
\end{equation*}
Since $\pi$ was chosen arbitrarily, this implies that $\psi^\varepsilon(t,x,y)  \geq \Val(t,x,y)$ and hence
\begin{equation*}
 U_\gamma\bigl(x + y\varphi_0(t)\bigr)e^{\gamma K(T-t)} = \lim_{\varepsilon\downarrow 0} \psi^\varepsilon(t,x,y)  \geq \Val(t,x,y).\qedhere
\end{equation*}
\end{proof}

\begin{oss}
	Another way to prove estimates on $\Val$ is to use financial arguments as follows:
	\begin{itemize}
		\item The lower bound corresponds to the value function of the problem without random endowments and can therefore be improved when random endowments are present;
		\item An alternative upper bound could be found by comparing the problem to the one of an artificial market model where the endowment can also be traded.
	\end{itemize}
\end{oss}

\subsection{Viscosity characterization of the value function}



The next step is to show that the value function is the unique viscosity solution of the HJB equation. We refer to \cite{crandall1992user} for the definition of (and all important results on) viscosity solutions of second order PDEs. While the fact that $v$ is a viscosity solution of the HJB equation is standard, the uniqueness result poses significant challenges in the case of $\gamma<0$ as the value function tends to $-\infty$ near the boundary of the state space. To handle this issue, we require a sufficiently strong comparison principle for the HJB equation.

Before turning to the comparison principle, let us record that the Hamiltonian
\begin{equation*}
 H : [0,T]\times\Dom\times\R\times\R^{2\times 2}\to\R,\qquad (t,x,y,p,M)\mapsto H(t,x,y,p,M)
\end{equation*}
given by
\begin{equation*}
 H(t,x,y,p,M) := \ancomment{\sup_{\pi\in\ctrl}}\Bigl\{ [(r+\sigma\theta\pi)x + y]p_1 + \mu_C(t)yp_2 + \frac{1}{2}\trace\bigl[\Sigma_\pi(t,x,y)\Sigma_\pi(t,x,y)^\top M\bigr] \Bigr\}
\end{equation*}
for $(t,x,y)\in[0,T]\times\Dom$, $p\in\R^2$ and $M\in\R^{2\times 2}$ with
\begin{equation*}
\Sigma_\pi(t,x,y) := \begin{pmatrix}  \sigma\pi x & 0 \\ \rho\sigma_C(t)y & \sqrt{1-\rho^2}\sigma_C(t)y  \end{pmatrix},
\end{equation*}
is both finite everywhere and continuous.
\ancomment{Note that, for this, it is crucial that $\ctrl$ is compact, since otherwise the Hamiltonian may diverge at some points of its domain, hence making the following viscosity characterization more involved.}
We also note that if $\varphi$ is a sufficiently smooth function, it holds that
\begin{equation*}
 H\bigl(t,x,y,\mathrm{D}\varphi(t,x,y),\mathrm{D}^2\varphi(t,x,y)\bigr) = \ancomment{\sup_{\pi\in\ctrl}}\bigl\{\L^\pi\varphi(t,x,y)\bigr\},
\end{equation*}
where $\mathrm{D}\varphi$ and $\mathrm{D}^2\varphi$ denote, respectively, the gradient and the Hessian of $\varphi$ with respect to the spatial variables $(x,y)$. One important consequence of the continuity of $H$ is that, by Lemma V.6.1 in \cite{fleming2006controlled}, we are allowed to apply the parabolic version of Ishii's lemma; see Theorem V.6.1 in \cite{fleming2006controlled}.

\begin{Teo}[Comparison principle]\label{theoremComparison}
	Let $u,v:[0,T]\times\Dom\to\R$ such that $u$ is an upper semi-continuous viscosity subsolution and $v$ is a lower semi-continuous viscosity supersolution of the HJB equation with
	\begin{equation}\label{eqnComparisonBounds}
	U_\gamma(x)e^{\gamma K(T-t)} \leq u(t,x,y),v(t,x,y)\leq U_\gamma\bigl(x + y\varphi_0(t)\bigr)e^{\gamma K(T-t)},\quad (t,x,y)\in[0,T]\times\Dom.
	\end{equation}
	Suppose furthermore that $u(T,\cdot) \leq v(T,\cdot)$ on $\Dom$. Then
	\begin{equation*}
	u(t,x,y) \leq v(t,x,y)\qquad\text{for all } (t,x,y)\in[0,T]\times\Dom.
	\end{equation*}
\end{Teo}

\begin{proof}
	Step 1: Setup of the proof and definitions. We argue by contradiction and suppose that there exists $(t^*,x^*,y^*)\in[0,T)\times\Dom$ such that
	\begin{equation}\label{eqnComparison_01}
	u(t^*,x^*,y^*) -  v(t^*,x^*,y^*) > 0.
	\end{equation}
    For $\delta>0$ and $n\in\N_0$, we introduce a function $\phi_n:[0,T]\times\Dom\times\Dom\to\R$ given by
	\begin{multline*}
	\phi_n(t,x,y,\bar x,\bar y) := u(t,x,y) - v(t,\bar x,\bar y) - \delta\psi(t,x,y) - \delta\psi(t,\bar x,\bar y) - \frac{n}{2}\Bigl[|x-\bar x|^2 + |y-\bar y|^2\Bigr].
	\end{multline*}
	Here, $\psi := \psi_{\bar\gamma,0,1}$ for some $\bar\gamma\in(0,1)$ with $\bar\gamma>\gamma$. We recall that $\psi:[0,T]\times\Dom\to\R$ is a strict classical supersolution of the HJB equation and takes the form
	\begin{equation*}
	\psi(t,x,y) = U_{\bar\gamma}\bigl(x + y\varphi_1(t)\bigr)e^{\bar \gamma K(T-t)} + \bigl(x^{-(1+|\gamma|)}+y^{-(1+|\gamma|)}\bigr)e^{(1+|\gamma|)\lambda(T-t)},\quad (t,x,y)\in [0,T]\times\Dom,
	\end{equation*}
	where $\varphi_1$ is strictly positive on $[0,T]$. Finally, we define a function $\phi:[0,T]\times\Dom\to\R$ by
    \begin{equation*}
    \phi(t,x,y) := u(t,x,y) - v(t,x,y) - 2\delta \psi(t,x,y),\qquad (t,x,y)\in[0,T]\times\Dom,
    \end{equation*}
    and we shall subsequently assume that $\delta>0$ is sufficiently small to guarantee that
	\begin{equation*}
	\phi(t^*,x^*,y^*) = u(t^*,x^*,y^*) -  v(t^*,x^*,y^*) - 2\delta \psi(t^*,x^*,y^*) > 0,
	\end{equation*}
	which is possible by~\eqref{eqnComparison_01}. 
	
	Step 2: Maximizers of $\phi_n$. For each $n\in\N_0$, we define
	\begin{equation*}
	M_n := \sup_{(t,x,y,\bar x,\bar y)\in[0,T]\times\Dom\times\Dom} \phi_n(t,x,y,\bar x,\bar y)\qquad\text{and}\qquad M := \sup_{(t,x,y)\in[0,T]\times\Dom} \phi(t,x,y).
	\end{equation*}
	Let us first observe that $M_n \geq M_{n+1} \geq M > 0$ for all $n\in\N_0$ as, clearly, $\phi_n$ is decreasing in $n$ (implying $M_n \geq M_{n+1}$) and
	\begin{equation*}
	M_n = \sup_{(t,x,y,\bar x,\bar y)\in[0,T]\times\Dom\times\Dom} \phi_n(t,x,y,\bar x,\bar y)
	\geq \sup_{(t,x,y)\in[0,T]\times\Dom} \phi(t,x,y) = M
	\geq \phi(t^*,x^*,y^*) > 0.
	\end{equation*}
	Moreover, using~\eqref{eqnComparisonBounds}, we note that for each $n\in\N_0$ and $(t,x,y,\bar x,\bar y)\in[0,T]\times\Dom\times\Dom$ we have
	\begin{align*}
	\phi_n(t,x,y,\bar x,\bar y)  &\leq u(t,x,y) - v(t,\bar x,\bar y) - \delta \psi(t,x,y) -\delta \psi(t,\bar x,\bar y)\\
	&\leq \Bigl[U_\gamma\bigl(x + y\varphi_0(t)\bigr)  - \delta U_{\bar\gamma}\bigl(x + y\varphi_1(t)\bigr)  - U_\gamma(\bar x) - \delta U_{\bar\gamma}\bigl(\bar x +\bar y \varphi_1(t)\bigr)\Bigr]e^{\bar\gamma K(T-t)}\\
	  &\hspace{1cm}  - \delta\Bigl[ x^{-(1+|\gamma|)} + y^{-(1+|\gamma|)} + \bar x^{-(1+|\gamma|)} + \bar y^{-(1+|\gamma|)}\Bigr]e^{(1+|\gamma|)\lambda (T-t)}.
	\end{align*}
Since $\bar\gamma>\max\{\gamma,0\}$ and $\varphi_1>0$, it follows that
	\begin{equation*}
	\sup_{n\in\N_0}\sup_{t\in[0,T]}\;\phi_n(t,x,y,\bar x,\bar y) \to -\infty\qquad\text{as } x \to \infty \text{ or } y \to \infty \text{ or }\bar x \to \infty \text{ or }\bar y \to \infty.
	\end{equation*}
	Similarly, we see that
	\begin{equation*}
	\sup_{n\in\N_0}\sup_{t\in[0,T]}\;\phi_n(t,x,y,\bar x,\bar y) \to -\infty\qquad\text{as } x \to 0 \text{ or } y \to 0 \text{ or }\bar x \to 0 \text{ or }\bar y \to 0.
	\end{equation*}
    In particular, as $\phi_0$ is upper semicontinuous, we find that
	\begin{equation*}
	\ancomment{F} := \bigl\{ (t,x,y,\bar x,\bar y)\in[0,T]\times\Dom\times\Dom : \phi_0 (t,x,y,\bar x,\bar y) \geq 0 \bigr\}\text{ is compact},
	\end{equation*}
	and, as $M_n > 0$ and $\phi_n \leq \phi_0$, any maximizing sequence for any $\phi_n$, $n\in\N_0$, is contained in \ancomment{$F$}. Moreover, upper semicontinuity of $\phi_n$ implies that we find \ancomment{$(t_n,x_n,y_n,\bar x_n,\bar y_n)\in F$} such that
	\begin{equation*}
	M_n = \sup_{(t,x,y,\bar x,\bar y)\in[0,T]\times\Dom\times\Dom} \phi_n(t,x,y,\bar x,\bar y) = \phi_n(t_n,x_n,y_n,\bar x_n,\bar y_n).
	\end{equation*}

		Step 3: Convergence of maximizers. As \ancomment{$F$} is compact and \ancomment{$(t_n,x_n,y_n,\bar x_n,\bar y_n)\in F$} for all $n\in\N_0$, after passing to a subsequence if necessary, it follows that
		\begin{equation*}
		\lim_{n\to\infty}(t_n,x_n,y_n,\bar x_n,\bar y_n) = (t_\infty,x_\infty,y_\infty,\bar x_\infty,\bar y_\infty)
		\end{equation*}
		for some \ancomment{$(t_\infty,x_\infty,y_\infty,\bar x_\infty,\bar y_\infty)\in F$}. Moreover, we note that $\psi,M_n\geq 0$ implies
		\begin{align}
		&\mathrel{\phantom{=}}\frac{n}{2}\Bigl(|x_n-\bar x_n|^2 + |y_n-\bar y_n|^2\Bigr)\notag\\
		&\hspace{2.5cm}= u(t_n,x_n,y_n) - v(t_n,\bar x_n,\bar y_n) - \delta\psi(t_n,x_n,y_n) - \delta\psi(t_n,\bar x_n,\bar y_n) - M_n\notag\\
		&\hspace{2.5cm}\leq \ancomment{\sup_{(t,x,y,\bar x,\bar y)\in F}} \bigl[u(t,x,y) - v(t,\bar x,\bar y)\bigr] < \infty.\label{eqn_proof_comparison_1}
		\end{align}
		For the finiteness of the last expression, we have used that $u$ and $-v$ are upper semicontinuous and finite-valued on the compact set \ancomment{$F$}. Since~\eqref{eqn_proof_comparison_1} is finite and does not depend on $n$, it follows that $(\bar x_\infty,\bar y_\infty) = (x_\infty,y_\infty)$. From this, using that $M_n\geq M$ and then upper semicontinuity of both $u$ and $-v$ and continuity of $\psi$, it follows that
		\begin{align*}
		0 & \leq \limsup_{n\to\infty} \frac{n}{2}\Bigl(|x_n-\bar x_n|^2 + |y_n-\bar y_n|^2\Bigr)\\
		& = \limsup_{n\to\infty}  \Bigl[u(t_n,x_n,y_n) - v(t_n,\bar x_n,\bar y_n) - \delta\psi(t_n,x_n,y_n) - \delta\psi(t_n,\bar x_n,\bar y_n) - M_n\Bigr]\\
		&\leq \limsup_{n\to\infty}  \Bigl[u(t_n,x_n,y_n) - v(t_n,\bar x_n,\bar y_n) - \delta\psi(t_n,x_n,y_n) - \delta\psi(t_n,\bar x_n,\bar y_n) - M\Bigr]\\
		&\leq u(t_\infty,x_\infty,y_\infty) - v(t_\infty,x_\infty,y_\infty) - \delta\psi(t_\infty,x_\infty,y_\infty) - \delta\psi(t_\infty,x_\infty,y_\infty) - M\\
		&=\phi(t_\infty,x_\infty,y_\infty) - M \leq 0.
		\end{align*}
		In particular, all inequalities must actually be equalities and the $\limsup$ can be replaced by a proper limit. We have therefore argued that
		\begin{equation*}
		\lim_{n\to\infty} \frac{n}{2}\Bigl(|x_n-\bar x_n|^2 + |y_n-\bar y_n|^2\Bigr) = 0\qquad\text{and}\qquad\lim_{n\to\infty} M_n = M = \phi(t_\infty,x_\infty,y_\infty)
		\end{equation*}
		as well as
		\begin{equation*}
		\lim_{n\to\infty} u(t_n,x_n,y_n) = u(t_\infty,x_\infty,y_\infty)\qquad\text{and}\qquad\lim_{n\to\infty} v(t,\bar x,\bar y) = v(t_\infty,x_\infty,y_\infty).
		\end{equation*}
		
		Step 4: Application of Ishii's lemma. Let us show that $(t_\infty,x_\infty,y_\infty)$ is not located on the boundary of the state space. \ancomment{As $F$ is a compact subset of $[0,T]\times\mathcal{O}\times\mathcal{O}$ and $\mathcal{O}=(0,\infty)\times (0,\infty)$}, we cannot have $x_\infty=0$ nor $y_\infty=0$. Moreover, if $t_\infty = T$, we use $\psi\geq 0$ and the assumption that $u(T,\cdot)\leq v(T,\cdot)$ on $\Dom$ to arrive at the contradiction
		\begin{align*}
		0 < M = \phi(T,x_\infty,y_\infty) \leq u(T,x_\infty,y_\infty) - v(T,x_\infty,y_\infty) \leq 0.
		\end{align*}
		Thus, as $t_\infty<T$, it follows that $t_n<T$ for all $n\in\N$ large enough and hence without loss of generality for all $n\in\N$.	We can therefore apply Theorem V.6.1 of \cite{fleming2006controlled} (Ishii's lemma) to obtain the existence of $s^n,\bar s^n\in\R$ and $M^n,\bar M^n\in\R^{2\times 2}$ symmetric such that $s^n - \bar s^n = 0$ and
		\begin{equation}\label{eqn_ishii_estimate}
		\begin{pmatrix}M^n & 0\\ 0 & -\bar M^n\end{pmatrix} \leq 3n\begin{pmatrix} \Id & -\Id\\ -\Id & \Id \end{pmatrix}\qquad\text{with}\qquad\Id := \begin{pmatrix}1 & 0\\ 0 & 1\end{pmatrix},
		\end{equation}
		and such that\footnote{Here, $\overline{\mathcal{J}}^{(1,2),+}w(t,x,y)$ and $\overline{\mathcal{J}}^{(1,2),-}w(t,x,y)$ denote the closures of the second-order parabolic super- and subjets of a function $w:[0,T]\times\Dom\to\R$ at $(t,x,y)$, respectively.}
		\begin{equation*}
		(q^n,p^n,P^n) \in\overline{\mathcal{J}}^{(1,2),+}u(t_n,x_n,y_n)\qquad\text{and}\qquad  (\bar q^n,\bar p^n,\bar P^n) \in\overline{\mathcal{J}}^{(1,2),-}v(t_n,\bar x_n,\bar y_n),
		\end{equation*}
		where
		\begin{align*}
		(q^n,p^n,P^n) &:= \left(s^n + \delta\psi_t(t_n,x_n,y_n),        n\begin{pmatrix}x_n-\bar x_n\\y_n-\bar y_n \end{pmatrix} + \delta\mathrm{D}\psi(t_n,x_n,y_n),                M^n + \delta\mathrm{D}^2\psi(t_n,x_n,y_n)\right),\\
		(\bar q^n,\bar p^n,\bar P^n) &:= \left(\bar s^n - \delta\psi_t(t_n,\bar x_n,\bar y_n),n\begin{pmatrix}x_n-\bar x_n\\y_n-\bar y_n \end{pmatrix} - \delta\mathrm{D}\psi(t_n,\bar x_n,\bar y_n),\bar M^n - \delta\mathrm{D}^2\psi(t_n,\bar x_n,\bar y_n)\right).
		\end{align*}
		
		Step 5: The contradiction. As $u$ and $v$ are viscosity sub- and supersolutions, it follows that
		\begin{equation*}
		-q^n - H\bigl(t_n,x_n,y_n,p^n,P^n\bigr) \leq 0\qquad\text{and}\qquad - \bar q^n - H\bigl(t_n,\bar x_n,\bar y_n,\bar p^n,\bar P^n\bigr) \geq 0.
		\end{equation*}
		Using the elementary inequality
		\begin{equation*}
		\sup\{a+b\} - \sup\{c-d\}  \leq \sup\{a-c\} + \sup\{b\} + \sup\{d\},
		\end{equation*}
		it follows that
		\begin{align}
		0 &\leq q^n - \bar q^n + H\bigl(t_n,x_n,y_n,p^n,P^n\bigr) - H\bigl(t_n,\bar x_n,\bar y_n,\bar p^n,\bar P^n\bigr)\label{eqnComparison_02}\\
		&\leq \ancomment{\sup_{\pi\in\ctrl}}\Bigl\{  (r+\sigma\theta\pi)n|x_n-\bar x_n|^2 + n(x_n-\bar x_n)(y_n-\bar y_n) + \mu_C(t_n)n|y_n-\bar y_n|^2\notag\\
		&\hspace{1.5cm} + \frac12\trace\Bigl[\Sigma_\pi(t_n,x_n,y_n)\Sigma_\pi(t_n,x_n,y_n)^\top M^n - \Sigma_\pi(t_n,\bar x_n,\bar y_n)\Sigma_\pi(t_n,\bar x_n,\bar y_n)^\top \bar M^n\Bigr]\Bigr\}\notag\\
		&\qquad + \delta\Bigl(\psi_t(t_n,\bar x_n,\bar y_n) + \ancomment{\sup_{\pi\in\ctrl}}\bigl\{\L^\pi\psi(t_n,\bar x_n,\bar y_n)\bigr\}\Bigr)\notag\\
		 &\qquad + \delta\Bigl(\psi_t(t_n,x_n,y_n) + \ancomment{\sup_{\pi\in\ctrl}}\bigl\{\L^\pi\psi(t_n,x_n,y_n)\bigr\}\Bigr).\notag
		\end{align}
		Using~\eqref{eqn_ishii_estimate}, a standard estimate shows that
		\begin{multline*}
		\frac12\trace\Bigl[\Sigma_\pi(t_n,x_n,y_n)\Sigma_\pi(t_n,x_n,y_n)^\top M^n - \Sigma_\pi(t_n,\bar x_n,\bar y_n)\Sigma_\pi(t_n,\bar x_n,\bar y_n)^\top \bar M^n\Bigr]\\
		 \leq \frac{3n}{2}\Bigl[\pi^2\sigma^2|x_n-\bar x_n|^2 + \sigma_C(t_n)^2|y_n-\bar y_n|^2\Bigr].
		\end{multline*}
		But then this and
		\begin{equation*}
		n(x_n-\bar x_n)(y_n-\bar y_n) \leq n\max\bigl\{|x_n-\bar x_n|^2,|y_n - \bar y_n|^2\bigr\} \leq n\bigl[|x_n-\bar x_n|^2 + |y_n - \bar y_n|^2\bigr]
		\end{equation*}
		allows us to continue to estimate~\eqref{eqnComparison_02} as follows:
		\begin{align*}
		0 &\leq \ancomment{\sup_{\pi\in\ctrl}}\Bigl\{(|r| + \sigma|\theta||\pi|)n|x_n-\bar x_n|^2 + n|x_n-\bar x_n|^2 + n|y_n - \bar y_n|^2\\
		&\hspace{2.5cm} +  |\mu_C(t_n)|n|y_n-\bar y_n|^2 + \frac{3}{2}\sigma^2\pi^2 n|x_n-\bar x_n|^2 + \frac{3}{2}\sigma_C(t_n)^2n|y_n-\bar y_n|^2 \Bigr\}\\
		&\quad + \delta\Bigl(\psi_t(t_n,\bar x_n,\bar y_n) + \ancomment{\sup_{\pi\in\ctrl}}\bigl\{\L^\pi\psi(t_n,\bar x_n,\bar y_n)\bigr\} + \psi_t(t_n,x_n,y_n) + \ancomment{\sup_{\pi\in\ctrl}}\bigl\{\L^\pi\psi(t_n,x_n,y_n)\bigr\}\Bigr)\\
		&\leq n C_x |x_n-\bar x_n|^2 + n C_y|y_n-\bar y_n|^2\\
		&\quad + \delta\Bigl(\psi_t(t_n,\bar x_n,\bar y_n) + \ancomment{\sup_{\pi\in\ctrl}}\bigl\{\L^\pi\psi(t_n,\bar x_n,\bar y_n)\bigr\} + \psi_t(t_n,x_n,y_n) + \ancomment{\sup_{\pi\in\ctrl}}\bigl\{\L^\pi\psi(t_n,x_n,y_n)\bigr\}\Bigr),
		\end{align*}
		where
		\begin{equation*}
		C_x := 1 + |r| + \sigma|\theta|\ancomment{\sup_{\pi\in\ctrl}}|\pi| + \frac{3}{2}\sigma^2\ancomment{\sup_{\pi\in\ctrl}}|\pi|^2
		\quad\text{and}\quad
		C_y := 1 + \sup_{t\in[0,T]}|\mu_C(t)| + \frac{3}{2}\sup_{t\in[0,T]}|\sigma_C(t)|^2.
		\end{equation*}
		Sending $n\to\infty$ and using that $\psi$ is a strict supersolution of the HJB equation therefore yields
		\begin{equation*}
		0 \leq 2\delta\Bigl(\psi_t(t_\infty,x_\infty,y_\infty) + \ancomment{\sup_{\pi\in\ctrl}}\bigl\{\L^\pi \psi(t_\infty,x_\infty,y_\infty)\bigr\}\Bigr) < 0,
		\end{equation*}
		which is the desired contradiction and hence concludes this proof.
\end{proof}

\begin{oss}
\ancomment{
The main technical challenge in extending the results in \cite{duffie1997hedging} to the case of risk aversion larger than one (i.e.\ $\gamma<0$) lies in the comparison principle. Indeed, for such values of $\gamma$, the utility function $U_\gamma$ explodes as $x\downarrow 0$, which requires very precise control of the viscosity sub-/supersolutions near that part of the boundary to establish a comparison principle. In our situation, this is achieved by Propostion~\ref{propTightBounds}, which allows us to pin down the behaviour of the value function near $x=0$.}
\end{oss}

With this comparison principle at hand, the stochastic Perron's method \cite{BayraktarSirbu2013} immediately implies that the value function $\Val$ is the unique continuous viscosity solution of the HJB equation.

\begin{Cor}[Viscosity characterization]
The value function $\Val$ is the unique viscosity solution of the HJB equation~\eqref{HJB_cont} in the class of continuous functions satisfying the terminal condition
\begin{equation*}
 \Val(T,x,y) = U_\gamma(x),\qquad (x,y)\in\Dom,
\end{equation*}
and the growth condition
\begin{equation*}
 U_\gamma(x)e^{\gamma K(T-t)} \leq \Val(t,x,y)\leq U_\gamma\bigl(x + y\varphi_0(t)\bigr)e^{K(T-t)},\quad (t,x,y)\in[0,T]\times\Dom.
\end{equation*}
\end{Cor}

\subsection{Regularity of the value function}

We now give sufficient conditions which guarantee that the value function is even a classical solution of the HJB equation. Related but different results can be found in \cite{federico2012impact} and \cite{federico2013utility}. The main obstacle to establishing regularity of the value function is the lack of uniform ellipticity of the HJB equation. However, the homotheticity property allows us to consider the transformation
\begin{equation}\label{eqnTrafo}
 \Val(t,x,y) = y^\gamma e^{-\beta t}w\bigl(t,\log x/y\bigr),\qquad\text{or, equivalently,}\qquad w(t,\zeta) := e^{\beta t}\Val(t,e^\zeta,1),
\end{equation}
where $\zeta := \log x/y$ and $\beta\in\R$. It turns out that $w$ solves a reduced-form HJB equation of the form
\begin{equation*}
	- w_t - \ancomment{\sup_{\pi\in\ctrl}}\Bigl\{ a(\pi,t)w_{\zeta\zeta}  + b(\pi,t,\zeta,w,w_\zeta)\Bigr\} = 0,\qquad (t,\zeta)\in(0,T)\times\R,
\end{equation*}
which is uniformly elliptic and admits a classical solution provided that $\mu_C$ and $\sigma_C$ are continuously differentiable. Moreover, observe that by passing to $w$ we have reduced the dimension of the state space by one, which is advantageous for numerical computations.


\begin{Teo}[Regularity]\label{ThmRegularity}
Assume that $\mu_C$ and $\sigma_C$ are continuously differentiable. Then $\Val\in C^{1,2}((0,T)\times\Dom)$. In particular, $\Val$ is a classical solution of the HJB equation.
\end{Teo}

\begin{proof}
Step 1: The transformed HJB equation. Let us consider the equation
\begin{equation}\label{eqnRegularSolution_01}
	- w_t - \ancomment{\sup_{\pi\in\ctrl}}\Bigl\{ a(\pi,t)w_{\zeta\zeta}  + b(\pi,t,\zeta,w,w_\zeta)\Bigr\} = 0,\qquad (t,\zeta)\in(0,T)\times\R,
\end{equation}
where
\begin{equation*}
 a:A\times[0,T]\to\R\qquad\text{and}\qquad b:A\times[0,T]\times\R\times\R\times\R\to\R
\end{equation*}
are given by
\begin{align*}
	a(\pi,t) &:= \frac{1}{2}\Bigl[\bigl(\pi\sigma - \rho \sigma_C(t)\bigr)^2 + (1-\rho^2)\sigma_C(t)^2\Bigr],\\
	b(\pi,t,\zeta,v,p) &:= \left(r - \mu_C(t) + \frac12(1-2\gamma)\sigma_C(t)^2 + \pi\sigma\theta - \frac12\pi^2\sigma^2+ \gamma\rho\sigma\sigma_C(t)\pi + e^{-\zeta}\right)p\\
	&\hspace{6cm} + \left(\gamma\mu_C(t) - \gamma\frac12(1-\gamma)\sigma_C(t)^2 +\beta\right)v,
\end{align*}
for \ancomment{$\pi\in\ctrl$}, $t\in[0,T]$, $\zeta,v,p\in\R$, and where the parameter $\beta\in\R$ is chosen such that
\begin{equation}\label{eqnRegularSolution_02}
	\beta < \inf_{t\in[0,T]} \Bigl[- \gamma \mu_C(t) + \frac12\gamma(1-\gamma)\sigma_C(t)^2\Bigr].
\end{equation}
Formally, Equation~\eqref{eqnRegularSolution_01} arises if we consider the transformation
\begin{equation*}
 w:[0,T]\times\R\to\R,\qquad (t,\zeta)\mapsto w(t,\zeta) := e^{\beta t}\Val(t,e^\zeta,1).
\end{equation*}
Now fix $N\in\N$. Then we claim that Equation~\eqref{eqnRegularSolution_01} admits a solution $w^N:(0,T]\times[-N,N]\to\R$ with $w^N\in C^{1,2}((0,T)\times(-N,N))$ satisfying the boundary and terminal conditions
\begin{equation}\label{eqnRegularSolution_03}
	w^N(t,\zeta) = e^{\beta t}\Val(t,e^\zeta,1),\qquad (t,\zeta)\in \Bigl(\{T\}\times[-N,N]\Bigr)\cup\Bigl((0,T)\times\{-N,N\}\Bigr).
\end{equation}
Once this is established, we can define a continuous function
\begin{equation*}
	v^N : (0,T]\times\overline{\Dom(N)} \to \R,\qquad (t,x,y)\mapsto v^N(t,x,y) := y^\gamma e^{-\beta t}w^N(t,\log x/y),
\end{equation*}
where $\overline{\Dom(N)}$ denotes the closure of the set
\begin{equation*}
	\Dom(N) := \bigl\{(x,y)\in\Dom : \log x/y \in (-N,N)\bigr\}.
\end{equation*}
Observe that, by the terminal/boundary condition~\eqref{eqnRegularSolution_03} and the homogeneity of $\Val$, we have
\begin{equation*}
 v^N(t,x,y) = y^\gamma e^{-\beta t}w^N(t,\log x/y) = y^\gamma \Val(t,x/y,1) = \Val(t,x,y)
\end{equation*}
whenever $t=T$ or $(x,y)\in\overline{\Dom(N)}\setminus\Dom(N)$. Moreover, since $w^N$ solves~\eqref{eqnRegularSolution_01} on $(0,T)\times(-N,N)$, a straightforward calculation shows that $v^N$ solves the original HJB equation, i.e.\
\begin{equation*}
-v^N_t(t,x,y) - \ancomment{\sup_{\pi\in\ctrl}}\bigl\{\L^\pi v^N(t,x,y)\bigr\} = 0,\qquad (t,x,y)\in(0,T)\times(-N,N).
\end{equation*}
But then $v^N$ also satisfies this equation in the sense of viscosity solutions and hence $v^N = \Val$ on $(0,T]\times\overline{\Dom(N)}$ by the uniqueness result Theorem V.8.1 in \cite{fleming2006controlled}. In particular, we find that $\Val\in \ancomment{C^{1,2}}((0,T)\times\Dom(N))$ for each $N\in\N$ and we conclude by sending $N\to\infty$.

Step 2: We are left with showing that Equation~\eqref{eqnRegularSolution_01} admits a classical solution $w^N$ on $(0,T)\times(-N,N)$ satisfying the boundary/terminal condition~\eqref{eqnRegularSolution_03}. For this, it is sufficient to verify the conditions in Theorem~A.8 in \cite{federico2012impact}; see also Theorem 3 in Section 6.4 of \cite{krylov1987nonlinear} for the original result.
\begin{enumerate}
	\item[(i)] For every \ancomment{$\pi\in\ctrl$}, it is clear that $a(\pi,\cdot)$ and $b(\pi,\cdot)$ are continuously differentiable and for each pair \ancomment{$(\pi,t)\in \ctrl\times[0,T]$} the function $b(\pi,t,\cdot)$ is twice continuously differentiable. Indeed, the non-zero derivatives are given by
	\begin{align*}
		a_t &= \sigma_C(t)\dot\sigma_C(t)  - \rho\sigma\dot\sigma_C(t)\pi,\\
		b_t &= \bigl(- \dot\mu_C(t) + (1-2\gamma)\sigma_C(t)\dot\sigma_C(t) + \gamma\rho\sigma\dot\sigma_C(t)\pi\bigr)p + \bigl(\gamma \dot\mu_C(t) - \gamma(1-\gamma)\sigma_C(t)\dot\sigma_C(t) +\beta\bigr)v,\\
		b_v &= \gamma \mu_C(t) - \frac12\gamma(1-\gamma)\sigma_C(t)^2 +\beta,\\
		b_p &= r - \mu_C(t) + \frac12(1-2\gamma)\sigma_C(t)^2 + \sigma\theta\pi - \frac12\sigma^2\pi^2+ \gamma\rho\sigma\sigma_C(t)\pi + e^{-\zeta},\\
		b_\zeta &= -b_{\zeta\zeta} =  -e^{-\zeta}p,\\
		b_{\zeta p} &=b_{p\zeta} -e^{-\zeta}.
	\end{align*}
	From this, we also see that $a_t,b_t$ as well as all second-order derivatives of $b$ with respect to $\zeta$ and $p$ are bounded on
	\begin{equation*}
		S_M := \bigl\{ \ancomment{(\pi,t,\zeta,v,p)\in \ctrl\times(0,T)\times(-N,N)\times \R\times\R} : |v| + |p| \leq M \bigr\}\quad\text{for all }M\in\N.
	\end{equation*}
	\item[(ii)] $a$ is uniformly elliptic on \ancomment{$\ctrl\times [0,T]$}, i.e.\
	\begin{equation*}
		0 < \frac{1}{2}(1-\rho)\inf_{t\in[0,T]} |\sigma_C(t)|^2  \leq a(\pi,t) \leq \ancomment{\sup_{(t,\pi)\in[0,T]\times\ctrl}} |a(\pi,t)| < \infty,\quad \ancomment{(\pi,t)\in \ctrl\times[0,T]}.
	\end{equation*}
	\item[(iii)] For all \ancomment{$(\pi,t,\zeta,v,p)\in \ctrl\times(0,T)\times(-N,N)\times\R\times\R$}, it holds that
	\begin{equation*}
		(1+|p|)\bigl|a_p(\pi,t)\bigr| + \bigl|a_v(\pi, t)\bigr| + \frac{\bigl|a_\zeta(\pi,t)\bigr|}{1+|p|} = 0.
	\end{equation*}
	Similarly, setting
	\begin{align*}
		C_1 &:= \ancomment{\sup_{(\pi,t,\zeta)\in \ctrl\times[0,T]\times[-N,N]}}\Bigl|r - \mu_C(t) + \frac12(1-2\gamma)\sigma_C(t)^2 + \sigma\theta\pi\\
		&\hspace{7cm} - \frac12\sigma^2\pi^2+ \gamma\rho\sigma\sigma_C(t)\pi + e^{-\zeta} \Bigr|,\\
		C_2 &:= \sup_{t\in[0,T]}\Bigl|\gamma \mu_C(t) - \frac12\gamma(1-\gamma)\sigma_C(t)^2 +\beta\Bigr|,
	\end{align*}
	it follows that
	\begin{multline*}
		(1+|p|)\bigl|b_p(\pi,t,\zeta,v,p)\bigr| + \bigl|b(\pi,t,\zeta,v,p)\bigr| + \bigl|b_v(\pi,t,\zeta,v,p)\bigr| + \frac{\bigl|b_\zeta(\pi,t,\zeta,v,p)\bigr|}{(1+|p|)}\\
		\leq (1+|p|)C_1 + |p|C_1 + |v|C_2 +C_2 +\frac{|p|}{1+|p|}e^{N} \leq h(v)\bigl(1 + |p|^2\bigr),
	\end{multline*}
	where $h:\R\to[0,\infty)$ is the continuous function given by
	\begin{equation*}
		h(v) := |v|C_2 + 3C_1 + C_2 + e^{N}, \qquad v\in\R.
	\end{equation*}
	\item[(iv)] Fix $Q>0$. Then, since $\beta$ is chosen to satisfy~\eqref{eqnRegularSolution_02}, there exists a constant $\delta>0$ such that
	\begin{align*}
		b(\pi,t,\zeta,-Q,0) &= -\left(\gamma \mu_C(t) - \frac12\gamma(1-\gamma)\sigma_C(t)^2 +\beta\right)Q \geq \delta>0,\\
		b(\pi,t,\zeta,Q,0) &= \left(\gamma \mu_C(t) - \frac12\gamma(1-\gamma)\sigma_C(t)^2 +\beta\right)Q \leq -\delta<0,
	\end{align*}
	for all \ancomment{$(\pi,t,\zeta)\in \ctrl\times(0,T)\times(-N,N)$}.
	\end{enumerate}
Under conditions (i) to (iv), Theorem 3 in Section 6.4 of \cite{krylov1987nonlinear} is applicable. This yields the existence of $w^N$ and the proof is complete. 
\end{proof}

\section{The optimal strategy and its asymptotic behavior}\label{sec:opt}

In this section, we construct an optimal strategy and study its properties as the ratio of wealth to endowment becomes large.

\subsection{Existence and characterization of  the optimal strategy}

A candidate optimal strategy is readily found by computing the maximizer in the HJB equation. Indeed, a straightforward calculation shows that the unique maximizer of the Hamiltonian $H$ is given by
\begin{equation}\label{eqnOptFeedback}
\hat\pi^*(t,x,y) := \Bigl(\Bigl[- \frac{\theta}{\sigma}\frac{\Val_x(t,x,y)}{x\Val_{xx}(t,x,y)}  - \frac{\rho\sigma_C(t)}{\sigma}\frac{y\Val_{xy}(t,x,y)}{x\Val_{xx}(t,x,y)}\Bigr]\vee \underline\pi\Bigr) \wedge \overline \pi,\quad(t,x,y)\in[0,T)\times\Dom.
\end{equation}
For any initial configuration $(t,x,y)$, a candidate optimal strategy \ancomment{$\pi^*\in\adm$} is defined on $(t,T)$ by\footnote{The strategy $\pi^*$ should of course be extended to an \ancomment{$\ctrl$-valued} process on $[0,T]$. Note, however, that the function $\hat\pi^*$ may not be well defined for $t\in\{0,T\}$. An argument as in Remark 3.1 (iv) in \cite{Touzi2012} shows that $\pi^*\mathds{1}_{[0,t)}$ can without loss of generality be taken independent of $\pi^*\mathds{1}_{[t,T]}$, e.g.\ equal to a constant value.}
\begin{equation*}
\pi^*_s := \hat\pi^*(s,A_s^*,c_s),\qquad s\in(t,T),
\end{equation*}
where $c := c^{t,y}$ and $A^*$ is the solution of
\begin{equation}\label{eqnOptWealth}
\ud A^*_s = \bigl[\bigl(r+\sigma\theta\hat\pi(s,A_s^*,c_s)\bigr)A^*_s + c_s\bigr]\ud s + \sigma \hat\pi(s,A_s^*,c_s) A^*_s\ud W^1_s,\quad s\in[t,T],\quad A^*_t = x.\quad
\end{equation}
We note, however, that is is unclear if the drift and diffusion coefficients
\begin{equation*}
(x,y) \mapsto \bigl(r+\sigma\theta\hat\pi(t,x,y)\bigr)x + y\qquad\text{and}\qquad (x,y)\mapsto\sigma \hat\pi(t,x,y)x
\end{equation*}
satisfy the necessary regularity to ensure the existence of a strong solution of \eqref{eqnOptWealth}. We therefore follow a different route: We first argue that for each initial configuration $(t,x,y)$, an optimal strategy exists and then show that it can be represented in feedback form via the function $\hat\pi^*$. \ancomment{This somewhat unusual route is an artifact of our assumption of $\ctrl$ being compact. In fact, we expect that allowing trading strategies to take values in the entire real line makes it possible to use similar arguments as in \cite{duffie1997hedging} to construct the optimal strategy in a more classical way.}

\begin{Prop}[Existence of optimizers]
Let $(t,x,y)\in[0,T)\times\Dom$. Then there exists an optimal trading strategy \ancomment{$\pi^* = \pi^*_{t,x,y}\in\adm$}, i.e.\
\begin{equation*}
	 \E\bigl[U_\gamma\bigl(A_T^{\pi^*} \bigr)\bigr] = \ancomment{\sup_{\pi\in\adm}} \E\bigl[U_\gamma\bigl(A_T^\pi \bigr)\bigr]  = \Val(t,x,y).
\end{equation*}
\end{Prop}

\begin{proof}
We employ a classical Komlos-type argument and show that suitable forward-convex combinations of a maximizing sequence for $\Val$ converge to an optimal strategy. Denote by \ancomment{$\{\hat \pi^n\}_{n\in\N}\subset\adm$} such a maximizing sequence for $\Val(t,x,y)$, i.e.\ a sequence of admissible trading strategies with
\begin{equation*}
 \Val(t,x,y) = \lim_{n\to\infty} \E\bigl[U_\gamma\bigl(\hat A_T^n \bigr)\bigr],\qquad \text{where }\hat A^n := A^{\hat \pi^n,t,x,y}.
\end{equation*}
For each $n\in\N$, let us denote by $\hat S^n := \hat \pi^n\hat A^n$ the wealth invested in the stock under the strategy $\hat \pi^n$. We may extend $\hat S^n$ to a process defined on $[0,T]$ by setting $\hat S^n_s := \hat \pi^n_s x$ for all $s\in[0,t]$. Now observe that, by uniform boundedness of $\{\hat \pi^n\}_{n\in\N}$ and the a priori estimate \eqref{eqn_apriori_growth_positive} in Lemma \ref{LemmaAPriori}, it holds that
\begin{equation*}
 \sup_{n\in\N} \E\Bigl[\int_0^T |\hat S^n_s|^2 \ud s\Bigr] < \infty,
\end{equation*}
i.e.\ the sequence $\{\hat S^n\}_{n\in\N}$ is bounded in the Hilbert space of progressively measurable and square-integrable processes. We can therefore apply Theorem 15.1.2 in \cite{DelbaenSchachermayer2006} to find $S^* = \{S^*_s\}_{s\in[0,T]}$ and a sequence $\{S^n\}_{n\in\N}$ with $S^n\in\mathrm{conv}\{\hat S^n,\hat S^{n+1},\ldots\}$ such that
\begin{equation*}
 \lim_{n\to\infty}\E\Bigl[\int_0^T \bigl|S^*_s - S^n_s\bigr|^2\ud s\Bigr] = 0.
\end{equation*}
Next, for each $n\in\N$, let us introduce processes $A^n = \{A^n_s\}_{s\in[0,T]}$ and $\pi^n = \{\pi^n_s\}_{s\in[0,T]}$ with $\pi^n := S^n / A^n$ and $A^n$ given as the unique solution of
\begin{equation*}
 \ud A^n_s = [rA^n_s + \sigma\theta S^n_s + c^{t,y}_s] \ud s + \sigma S^n_s\ud W^1_s,\qquad s\in[t,T],
 \end{equation*}
with $A^n = x$ on $[0,t]$. Observe that $A^n$ is the wealth process corresponding to the trading strategy $\pi^n$. Moreover, since $S^n\in\mathrm{conv}\{\hat S^n,\hat S^{n+1},\ldots\}$, there exist $N_n\in\N$, indices $k^n_1,\ldots,k^n_{N_n}\in\{n,n+1,\ldots\}$ and convex weights $\lambda_1^n,\ldots,\lambda^n_{N_n}$ such that $S^n = \sum_{i=1}^{N_n} \lambda^n_i \hat S^{k^n_i}$.
With this, we see that
\begin{align*}
 \ud \sum_{i=1}^{N_n} \lambda^n_i \hat A^{k^n_i}_s &= \sum_{i=1}^{N_n} \lambda^n_i\Bigl[[r\hat A^{k^n_i}_s +\sigma\theta \hat S^{k^n_i}_s + c^{t,y}_s]\ud s + \sigma \hat S^{k^n_i}_s\ud W^1_s \Bigr]\\
  &= \Bigl[r\sum_{i=1}^{N_n} \lambda^n_i\hat A^{k^n_i}_s +\sigma\theta S^n_s + c^{t,y}_s\Bigr]\ud s + \sigma S^n_s\ud W^1_s, & s\in[t,T],
\end{align*}
from which we conclude $A^n = \sum_{i=1}^{N_n} \lambda^n_i \hat A^{k^n_i}$ by uniqueness of solutions of linear SDEs. But then
\begin{equation*}
 \pi^n = \frac{S^n}{A^n} = \sum_{k=1}^{N_n} \frac{\lambda^n_i \hat A^{k^n_i}}{A^n}\hat\pi^{k^n_i},
\end{equation*}
i.e.\ $\pi^n$ is a (dynamic) convex combination of $\pi^{k^n_1},\ldots,\pi^{k^n_{N_n}}$ and therefore \ancomment{$\ctrl$-valued} and admissible. Next, consider the wealth process $A^* = \{A^*_s\}_{s\in[0,T]}$ given by
\begin{equation*}
\ud A^*_s = [rA^*_s + \sigma\theta S^*_s + c^{t,y}_s] \ud s + \sigma S^*_s\ud W^1_s,\qquad s\in[t,T],
\end{equation*}
with $A^* = x$ on $[0,t]$ and observe that, using the dynamics of $A^n$ and $A^*$, Jensen's inequality, and the It\^{o} isometry, there exists a constant $C>0$ such that
\begin{align*}
 \E\bigl[|A^*_s - A^n_s|^2\bigr] &\leq C\E\Bigl[\int_t^s |A^*_u - A^n_u|^2\ud u\Bigr] + C\E\Bigl[\int_t^s  |S^*_u - S^n_u|^2 \ud u \Bigr]\\
   &\leq C\int_t^s \E\bigl[|A^*_u - A^n_u|^2\bigr]\ud u + C\E\Bigl[\int_0^T  |S^*_u - S^n_u|^2 \ud u \Bigr], & s\in[t,T].
\end{align*}
Gronwall's inequality therefore yields
\begin{equation*}
 \lim_{n\to\infty}\E\bigl[|A^*_s - A^n_s|^2\bigr]  \leq \lim_{n\to\infty}C\E\Bigl[\int_0^T  |S^*_u - S^n_u|^2 \ud u \Bigr] e^{C(T-t)} = 0,\qquad s\in[t,T].
\end{equation*}
Setting $\pi^* := S^*/A^*$, using the convergence of $(S^n,A^n)$ to $(S^*,A^*)$, it follows that $\pi^*$ is \ancomment{$\ctrl$-valued} $\ud t\otimes\P$ almost everywhere (hence without loss of generality everywhere after possibly redefining $\pi^*$ on a nullset) and hence admissible. Clearly, $A^*$ is the wealth process corresponding to $\pi^*$. Moreover, since $U_\gamma(A^n_T)^2 = |A^n_T|^{2\gamma}/|\gamma|^2$, it follows from Lemma~\ref{LemmaAPriori} that $\{U_\gamma(A^n_T)\}_{n\in\N}$ is bounded in the space of square-integrable random variables and thus uniformly integrable. With this and using the concavity of $U_\gamma$, we hence conclude that
\begin{equation*}
 \E\bigl[U_\gamma\bigl(A^*_T\bigr)\bigr] = \lim_{n\to\infty} \E\bigl[U_\gamma\bigl(A^n_T\bigr)\bigr]  \geq \lim_{n\to\infty} \sum_{i=1}^{N_n}\lambda^n_i \E\bigl[U_\gamma\bigl(\hat A^{k^n_i}_T\bigr)\bigr] = \Val(t,x,y).\qedhere
\end{equation*}
\end{proof}

Next, let us proceed to show that any optimal strategy must necessarily be given in feedback form via the function $\hat\pi^*$ defined in~\eqref{eqnOptFeedback}. 

\begin{Teo}[Optimal strategy in feedback form]
	Fix $(t,x,y)\in[0,T)\times\Dom$, let \ancomment{$\pi\in\adm$} be an arbitrary strategy, and suppose that $\Val\in C^{1,2}((0,T)\times\Dom)$. Then $\pi$ is optimal if and only if
	\begin{equation}\label{eqnOptStrat}
	 \pi_s(\omega) = \hat\pi^*\bigl(s,A_s(\omega),c_s(\omega)\bigr)\qquad\text{for }\P\otimes\ud s\text{-almost every }(\omega,s)\in\Omega\times\bigl([t,T]\cap(0,T)\bigr),
	\end{equation}
	where $A := A^{\pi,t,x,y}$ and $c := c^{t,y}$.
\end{Teo}

\begin{proof}
Step 1: Suppose that \ancomment{$\pi\in\adm$} is an arbitrary strategy satisfying \eqref{eqnOptStrat}. Using that $\Val$ solves the HJB equation and $\hat\pi^*$ is a pointwise maximizer of the supremum in the HJB equation, an application of It\^{o}'s formula shows that
\begin{align*}
\Val\bigl(s,A_s,c_s\bigr) &= \Val(t,x,y) + \int_t^s \Val_t(u,A_u,c_u) + \L^{\pi_u}\Val(u,A_u,c_u)\ud u\\
&\hspace{2.5cm}+  \int_t^s \sigma\pi_uA_u \Val_x(u,A_u,c_u)  \ud W^1_u + \int_t^s \sigma_C(u)c_u\Val_y(u,A_u,c_u) \ud W^C_u\\
&= \Val(t,x,y) +  \int_t^s \sigma\pi_uA_u \Val_x(u,A_u,c_u)  \ud W^1_u + \int_t^s \sigma_C(u)c_u\Val_y(u,A_u,c_u) \ud W^C_u
\end{align*}
for all $s\in[t,T]$, i.e.\ \ancomment{$\Val(\cdot,A,c)$} is a local martingale. But since by~\eqref{eqnValPowerGrowth} and Lemma~\ref{LemmaAPriori}
\begin{equation}\label{eqnLocalMartingaleHonest}
\E\Bigl[\sup_{s\in[t,T]} \bigl| \Val\bigl(s,A_s,c_s\bigr) \bigr|\Bigr] \leq C\E\Bigl[1 + \sup_{s\in[t,T]}|A_s|^{-\gamma^-} + \sup_{s\in[t,T]}|A_s|^{\gamma^+} + \sup_{s\in[t,T]}|c_s|^{\gamma^+}\Bigr] < \infty,
\end{equation}
it follows that $\Val(\cdot,A,c)$ is an honest martingale and hence
\begin{equation*}
\Val(t,x,y) = \E\bigl[\Val\bigl(T,A_T,c_T\bigr)\bigr] = \E\bigl[U_\gamma\bigl(A_T\bigr)\bigr],
\end{equation*}
i.e.\ $\pi$ is optimal.

Step 2: Suppose that \ancomment{$\pi\in\adm$} is optimal, i.e.\
\begin{equation}\label{eqnMartingale}
 \Val(t,x,y) = \E\bigl[U_\gamma\bigl(A_T\bigr)\bigr] = \E\bigl[\Val\bigl(T,A_T,c_T\bigr)\bigr].
\end{equation}
An application of It\^{o}'s formula and the fact that $\Val$ satisfies the HJB equation implies
\begin{align}
\Val\bigl(s,A_s,c_s\bigr) &= \Val(t,x,y) + \int_t^s \Val_t(u,A_u,c_u) + \L^{\pi_u}\Val(u,A_u,c_u)\ud u\notag\\
&\hspace{2.5cm}+  \int_t^s \sigma\pi_uA_u \Val_x(u,A_u,c_u)  \ud W^1_u + \int_t^s \sigma_C(u)c_u\Val_y(u,A_u,c_u) \ud W^C_u\notag\\
&\leq \Val(t,x,y) +  \int_t^s \sigma\pi_uA_u \Val_x(u,A_u,c_u)  \ud W^1_u + \int_t^s \sigma_C(u)c_u\Val_y(u,A_u,c_u) \ud W^C_u\label{eqnSupermartingale}
\end{align}
for all $s\in[t,T]$. But as \eqref{eqnLocalMartingaleHonest} is also valid in this situation, it follows that $\Val(\cdot,A,c)$ is a super-martingale and thus by \eqref{eqnMartingale} an honest martingale. Thus we must have equality in \eqref{eqnSupermartingale}, i.e.\
\begin{equation*}
 \Val_t(s,A_s,c_s) + \L^{\pi_s}\Val(s,A_s,c_s) = 0 = \Val_t(s,A_s,c_s) + \ancomment{\sup_{\pi\in\ctrl}} \bigl\{ \L^\pi\Val(s,A_s,c_s) \bigr\}
\end{equation*}
$\P\otimes\ud s$-almost everywhere on $\Omega\times\bigl([t,T]\cap(0,T)\bigr)$. Now the supremum on the right hand side has a unique maximizer given by the function $\hat\pi^*$ and thus
\begin{equation*}
\pi_s(\omega) = \hat\pi^*\bigl(s,A_s(\omega),c_s(\omega)\bigr)\qquad\text{for }\P\otimes\ud s\text{-almost every }(\omega,s)\in\Omega\times\bigl([t,T]\cap(0,T)\bigr)
\end{equation*}
and the proof is complete.
\end{proof}

\subsection{Asymptotic behavior of the optimal strategy}

In this section, we examine the asymptotic behavior of the value function and the optimal policy as the initial capital converges to infinity. We will see that for this model a sort of ``turnpike property'' holds (see e.g. \cite{zaslavski2006turnpike}) in the sense that for $x\gg y$ (more precisely, as $x/y\to\infty$ becomes large) the optimal policy approaches the Merton fraction $\pi_M$ of investing a constant proportion of wealth in the risky asset.

To make this statement precise, we subsequently write
\begin{equation*}
 f(z) \prop{z\to\infty} g(z)\qquad\text{if and only if}\qquad \lim_{z\to\infty} \frac{f(z)}{g(z)} = 1
\end{equation*}
whenever $f,g:\R\to\R$. Our aim in this section is hence to show that $\hat\pi^*(t,x,y) \prop{x/y\to\infty} \pi_M$ for all $t\in(0,T)$, where $\hat\pi^*$ is the feedback function defined in \eqref{eqnOptFeedback}. To prove this, we first recall the reduced value function $\ValRed$ defined in Lemma \ref{gen_homog_red} and observe that
\begin{align*}
 \Val_x(t,x,y) &= y^{\gamma-1}\ValRed_z(t,z), & \Val_{xx}(t,x,y) &= y^{\gamma-2}\ValRed_{zz}(t,z),\\
 \Val_{xy}(t,x,y) &= -y^{\gamma-2}\bigl[(1-\gamma)\ValRed_z(t,z) + z \ValRed_{zz}(t,z)\bigr].
\end{align*}
From this, it follows that $\hat\pi^*$ can be rewritten as
\begin{align}
\label{eq:piz}\hat\pi^*(t,z) =& \Bigl(\Bigl[- \frac{\theta}{\sigma}\frac{\ValRed_z(t,z)}{z\ValRed_{zz}(t,z)}  + \frac{(1-\gamma)\rho\sigma_C(t)}{\sigma}\frac{\ValRed_z(t,z)}{z\ValRed_{zz}(t,z)}  + \frac{\rho\sigma_C(t)}{\sigma}\Bigr]\vee \underline\pi\Bigr) \wedge \overline \pi,\\&(t,z)\in[0,T)\times(0,\infty).\nonumber
\end{align}

\begin{Teo}[Asymptotics for the optimal strategy]
For any $t\in(0,T)$, we have
\begin{align*}
 \ValRed(t,z) &\prop{z\to\infty} \frac{1}{\gamma}z^\gamma e^{\gamma K(T-t)}, & \ValRed_z(t,z) &\prop{z\to\infty} z^{\gamma-1} e^{\gamma K(T-t)},\\
 \ValRed_{zz}(t,z) &\prop{z\to\infty} -(1-\gamma)z^{\gamma-2} e^{\gamma K(T-t)},
\end{align*}
from which it follows that
\begin{equation*}
 \hat\pi^*(t,z) \prop{z\to\infty} \pi_M = \frac{\theta}{(1-\gamma)\sigma}.
\end{equation*}
\end{Teo}

\begin{proof}
	The statements on the asymptotic behavior of $\ValRed$ and its derivatives follow from the bounds on $\Val$ derived in Proposition \ref{propTightBounds}. Indeed, the bounds on $\Val$ and homogeneity of $U_\gamma$ imply that
	\begin{equation*}
	 \frac{1}{\gamma}z^\gamma e^{\gamma K(T-t)} \leq \ValRed(t,z) \leq \frac{1}{\gamma}\bigl(z + \varphi_0(t)\bigr)^\gamma e^{\gamma K(T-t)},\qquad (t,z)\in [0,T]\times(0,\infty).
	\end{equation*}
	From this and using monotonicity and concavity of all functions involved, we see right away that
	\begin{align*}
	\ValRed(t,z) &\prop{z\to\infty} \frac{1}{\gamma}z^\gamma e^{\gamma K(T-t)}, & \ValRed_z(t,z) &\prop{z\to\infty} z^{\gamma-1} e^{\gamma K(T-t)},\\
	\ValRed_{zz}(t,z) &\prop{z\to\infty} -(1-\gamma)z^{\gamma-2} e^{\gamma K(T-t)},
	\end{align*}
	But then
	\begin{equation*}
	 \lim_{z\to\infty} \frac{\ValRed_z(t,z)}{z\ValRed_{zz}(t,z)} = \lim_{z\to\infty} \frac{z^{\gamma-1} e^{\gamma K(T-t)}}{-(1-\gamma)z^{\gamma-1} e^{\gamma K(T-t)}} = -\frac{1}{1-\gamma},
	\end{equation*}
	and it follows that
	\begin{equation*}
	 \lim_{z\to\infty} \Bigl[- \frac{\theta}{\sigma}\frac{\ValRed_z(t,z)}{z\ValRed_{zz}(t,z)}  + \frac{(1-\gamma)\rho\sigma_C(t)}{\sigma}\frac{\ValRed_z(t,z)}{z\ValRed_{zz}(t,z)}  + \frac{\rho\sigma_C(t)}{\sigma}\Bigr] = \frac{\theta}{(1-\gamma)\sigma} = \pi_M.
	\end{equation*}
	Since \ancomment{$\pi_M\in\ctrl$}, we see that $\lim_{z\to\infty}\hat\pi^*(t,z) = \pi_M$ and the proof is complete.
\end{proof}

\section{Numerical illustration}\label{sec:num}
\ancomment{The strength of our approach is that we obtain a formula for the optimal strategy that is explicit in terms of the solution of the boundary value problem for the value function which in turn is highly amenable to numerical methods (for the plots shown in this section we implemented the finite difference methods of \cite{wang2008maximal} in Matlab). This allows one to consider questions of high economic relevance. Below we briefly look at two of them, viz.\ the optimal equity holdings of young and old investors in their pension fund and later on whether the common wisdom that ``the closer one is to retirement the lower the equity holdings should be'' is indeed universally true. Throughout this section we take $\mu_C,\sigma_C$ constant over time.}

\begin{figure}[t]
    \centering
	\includegraphics[width=\textwidth]{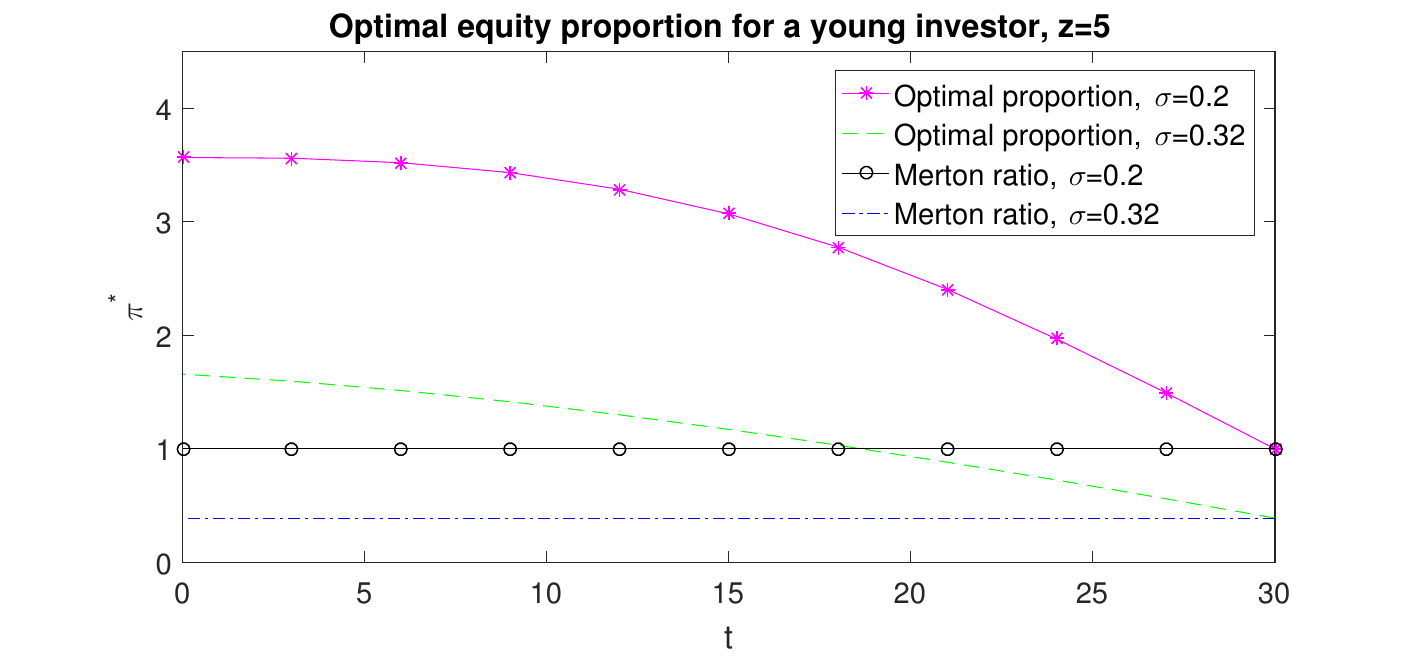}
	\caption{Optimal equity proportion for a young investor.\\
		\small{ Here {\bf z=5,\ T=30} years, $\rho=0.25$ and $\sigma = 0.2,0.32$, $\mu = 0.04$, $\sigma_C = 0.13$, $\mu_C = 0.02$, $r=0.02$, $\gamma = 0.5$.}}
		\label{young}
\end{figure}

\begin{figure}[t]
	\centering
	\includegraphics[width=\textwidth]{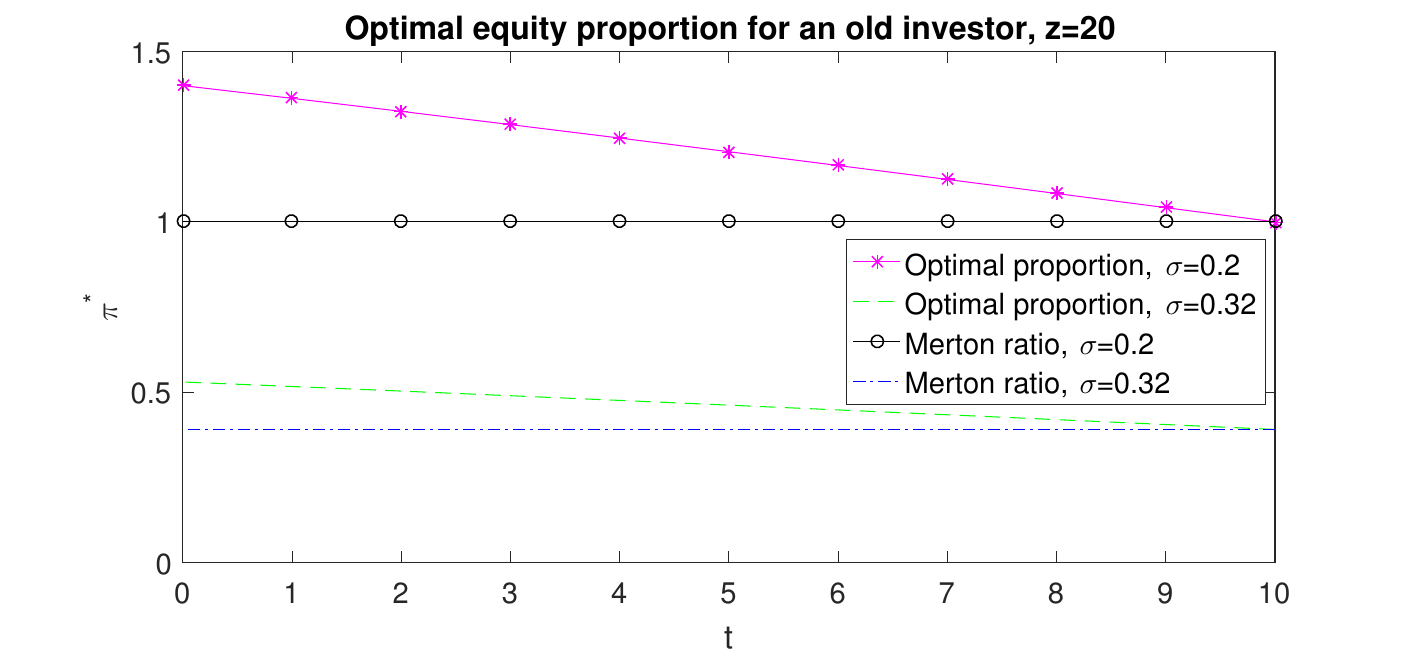}
	\caption{Optimal equity proportion for an old investor.\\
		\small{ Here {\bf z=20,\ T=10} years, $\rho=0.25$ and $\sigma = 0.2, 0.32$, $\mu = 0.04$, $\sigma_C = 0.13$, $\mu_C = 0.02$, $r=0.02$, $\gamma = 0.5$.}}
		\label{old}
\end{figure}

 \ancomment{We now consider the optimal equity holdings as a function of time for an old and a young investor. We distinguish between the two investors by choosing a different initial value of the ratio between current wealth and income $z=x/y$ and a different investment horizon $T$. A higher $z$ means a higher initial capital compared to the initial contribution. Note that the initial optimal investment strategy therefore always depends on the ratio of wealth over income. The young investor still needs to work for another 30 years until retirement and starts with a lower initial $z=5$, while the old investor only needs to work for another 10 years and has an initial $z=20$. The resulting optimal equity holdings are plotted in Figures \ref{young} and \ref{old} respectively. To gain a better intuition, Merton's constant-mix portfolios are provided in the two graphics as well. }
 

\ancomment{Comparing the solid curves in these two graphics, we observe the following: a) while Merton's portfolio is a constant-mix one, which does not depend on time, the equity holdings resulting from random endowments are decreasing in time, demonstrating a so-called glide path. The closer the individual investors move to retirement, the less will be invested in the risky asset. b) Young investors have a longer time to work and have not had much time to accumulate wealth. The ability to work (human capital) is therefore their largest asset. Older investors have already converted most of their human capital to financial capital.  In this sense, young investors can borrow from their future income to invest more in the risky asset, which leads to a substantially higher equity holding of the young investor. c) A higher volatility (keeping the drift fixed) makes the equity investment less interesting, which subsequently lowers the optimal equity holding. }

\begin{figure}[t]
\begin{center}
\includegraphics[width=\textwidth]{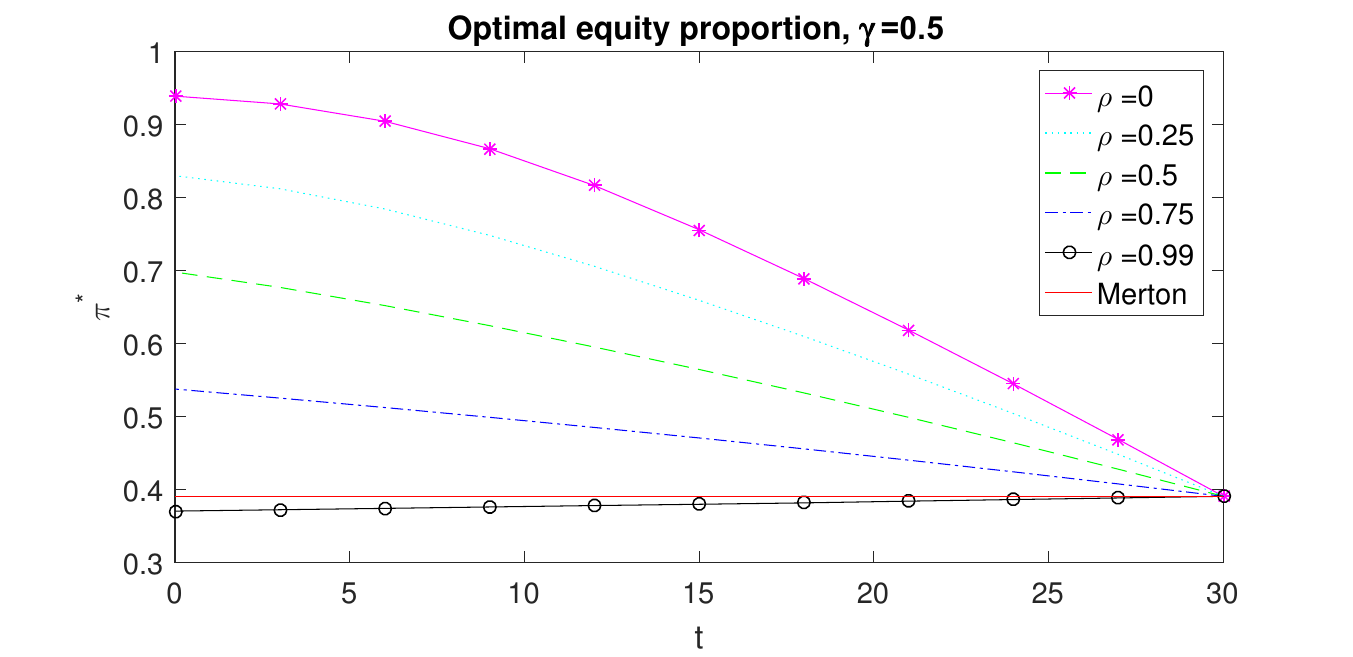}
\end{center}\caption{Optimal equity proportion  for different values of the correlation  $\rho$ and positive $\gamma$.\\
\small{ Parameters: $T=30$ years, $\sigma = 0.32$, $\mu = 0.04$, $\sigma_C = 0.13$, $\mu_C = 0.02$, $r=0.02$, $\gamma = 0.5$, $z=15$.}}\label{gamma_pos}
\end{figure}
\begin{figure}[t]
\begin{center}
\includegraphics[width=\textwidth]{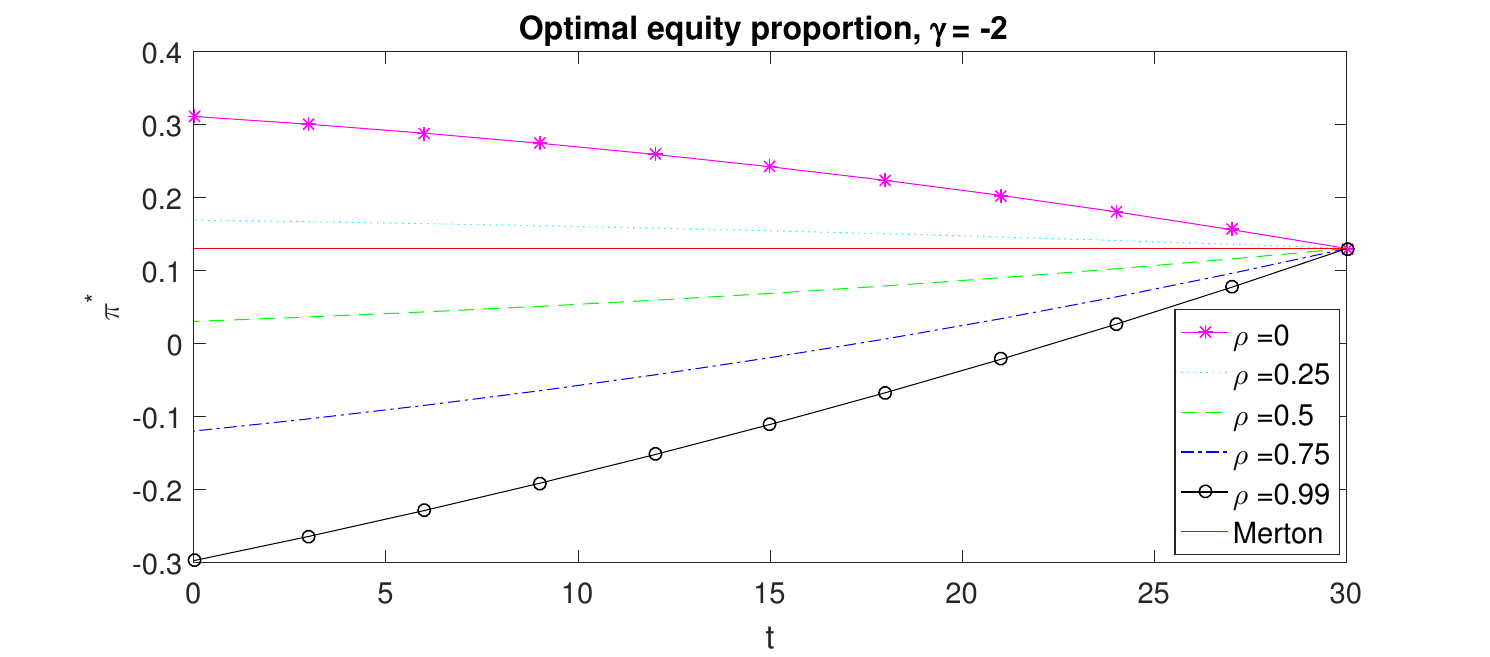}
\end{center}\caption{Optimal equity proportion  for different values of the correlation  $\rho$ and negative $\gamma$.\\
\small{ Parameters: $T=30$ years,  $\sigma = 0.32$, $\mu = 0.06$, $\sigma_C = 0.13$, $\mu_C = 0.02$, $r=0.02$, $\gamma = -2$, $z=15$.}}\label{corrrra3}
\end{figure}

\ancomment{In the above examples the common wisdom ``the closer you are to retirement the less risky you should invest'', which is also behind many products sold to small private investors, seems to be true and our intuitive explanation is that a young investor borrows from her future income enabling her to go more risky. It is realistic that the income and the stock market are correlated, in most cases (except for e.g. liquidators) positively. This implies that future income changes can at least be partially hedged by investing in the stock market. In the case of a positive correlation this should imply that a risk averse investor should go short in the stock market to hedge her income. This effect should be more pronounced the more risk averse an investor is. Figures \ref{gamma_pos} and \ref{corrrra3} depicting the optimal equity proportions for different values of $\rho$ as a function of time for $\gamma=0.5$ and the more risk averse $\gamma=-2$ show that our model can reproduce these effects. The higher the correlation $\rho$ is, the lower (ceteris paribus) is the investment in the stock market in order to also hedge the future income changes.  Note that for $\gamma=-2$ we have also taken a higher $\mu$ which makes an investment in the stock market more attractive. So for the same $\mu$ as in Figure \ref{gamma_pos} the optimal equity proportions held would be even lower. Interestingly, it turns out that the hedging effect can even be dominant implying that the closer one is to $T$ the higher is the optimal investment in the risky asset. This clearly shows that the mentioned common wisdom is far from being universally true: the more risk averse investors are, the less often it is true. }

\ancomment{To characterize exactly when we have an optimal equity proportion above or below the Merton ratio seems to be a very challenging question. What we can establish is the following.\begin{Prop}\label{critical}
Assume that $\sigma_C$ is constant over time and $\mu>r$. If  $1-\gamma>(\mu-r)/(\sigma\sigma_C)$, then for the correlation $\rho$ of the Brownian motions being equal to
\begin{equation}\label{critical_corr}
\rho^* = \frac{\mu-r}{\sigma\sigma_C(1-\gamma)}\ge0,
\end{equation}
the corresponding optimal strategy is constant and coincides with the Merton ratio, namely
\begin{equation*}
\pi^*(t,z)=\pi_M=\frac{\mu-r}{\sigma^2(1-\gamma)},\qquad (t,z)\in[0,T)\times (0,\infty).
\end{equation*}
\end{Prop}
\begin{proof} We look for a sufficient condition on the correlation such that the optimal policy in \eqref{eq:piz} coincides with the Merton ratio.
Define, for $z= x/y$,
\begin{equation*}
\delta(t,z) := -\frac{\ValRed_{zz}(t,z)}{\ValRed_z(t,z)} z - (1-\gamma).
\end{equation*}
Suppressing the arguments of $\delta$, we now look for a $\rho^*$ such that
\begin{equation*}
\frac{\mu-r}{\sigma^2(1-\gamma)} = \frac{\mu-r}{\sigma^2(1-\gamma + \delta)} - \rho^*\frac{\sigma_C}{\sigma}\left(\frac{1-\gamma}{1-\gamma + \delta} - 1\right).
\end{equation*}
Rearranging the terms, this holds if and only if
\begin{equation*}
\frac{\mu-r}{\sigma}\left(\frac{1}{1-\gamma}-\frac{1}{1-\gamma + \delta} \right) = - \rho^*\sigma_C(1-\gamma)\left(\frac{1}{1-\gamma + \delta} - \frac1{1-\gamma}\right).
\end{equation*}
Clearly, this equation is satisfied for $\rho^*=\frac{\mu-r}{\sigma\sigma_C(1-\gamma)}$.
The fact that $\rho^*$ is strictly positive follows from Equation \eqref{critical_corr} since $\mu>r$. Moreover, $\rho^*$ is a correlation coefficient thanks to the assumptions on the parameters.
\end{proof}}
\ancomment{Combining this with our numerical results and assuming that $\pi_M$ lies in the interior of $\ctrl$, it seems reasonable to conjecture that:
\begin{itemize}
\item if $\rho>\rho^*$, then $\pi^*(t,z)<\pi_M $ for all $(t,z)\in[0,T)\times (0,\infty)$,
\item if $\rho<\rho^*$, then $\pi^*(t,z)>\pi_M$ for all $(t,z)\in[0,T)\times (0,\infty)$.
\end{itemize}
This would also imply that for a sufficiently small relative risk aversion parameter (that is, $\gamma$ sufficiently large), all values of the correlation should yield optimal strategies lying above the Merton ratio. In the set-up of Figure \ref{gamma_pos}, the critical value is  $\rho^*\approx 0.9615$ and $\rho^*\approx0.3205$ for the parameter constellation of Figure \ref{corrrra3}.
}

\section{Conclusion}\label{sec:conclusion}
We consider an optimal asset allocation problem in an incomplete market, where exogenous stochastic endowments flow continuously into the portfolio according to a time-inhomogeneous geometric Brownian motion. We analyze the viscosity solution of the HJB PDE, reduce its dimension, and prove that the optimal strategy can be recovered from the optimal policy of a reduced problem.

We are also able to describe the asymptotic behavior of the value function, and the strategy when the initial wealth goes to infinity. \ancomment{We illustrate that our results open the door for precise numerical studies and briefly explain some economic insights to be gained.}  

\section*{Acknowledgements}
 \ancomment{The authors thank the editors as well as the anonymous referees for their insightful and constructive comments which  improved the paper significantly. They are very grateful to Nils S\o rensen for preparing the plots included in this paper. Carla Mereu gratefully acknowledges financial support from the Graduiertenkolleg 1100 at Ulm University, funded by the DFG (Deutsche Forschungsgemeinschaft).}

\bibliographystyle{siam}
\bibliography{ins_bibliography}

\begin{thebibliography}{10}

\bibitem{azar2010bounds}
{\sc S.~A. Azar}, {\em Bounds to the coefficient of relative risk aversion},
  Banking and Finance Letters, 2 (2010), pp.~391--398.

\bibitem{barsky1997preference}
{\sc R.~B. Barsky, F.~T. Juster, M.~S. Kimball, and M.~D. Shapiro}, {\em
  Preference parameters and behavioral heterogeneity: An experimental approach
  in the health and retirement study}, The Quarterly Journal of Economics, 112
  (1997), pp.~537--579.

\bibitem{BayraktarSirbu2013}
{\sc E.~Bayraktar and M.~S{\^{\i}}rbu}, {\em Stochastic {P}erron's method for
  {H}amilton-{J}acobi-{B}ellman equations}, SIAM Journal on Control and
  Optimization, 51 (2013), pp.~4274--4294.

\bibitem{BelakEtAl2015}
{\sc C.~Belak, O.~Menkens, and J.~Sass}, {\em On the uniqueness of unbounded
  viscosity solutions arising in an optimal terminal wealth problem with
  transaction costs}, SIAM Journal on Control and Optimization, 53 (2015),
  pp.~2878--2897.

\bibitem{Bick2013}
{\sc B.~Bick, H.~Kraft, and C.~Munk}, {\em Solving constrained
  consumption-investment problems by simulation of artificial market
  strategies}, Management Science, 59 (2013), pp.~483--503.

\bibitem{bosserhoff2021investment}
{\sc F.~Bosserhoff, A.~Chen, N.~S{\o}rensen, and M.~Stadje}, {\em On the
  investment strategies in occupational pension plans}, forthcoming in
  Quantitative Finance,  (2021).

\bibitem{broeders2010pension}
{\sc D.~Broeders and A.~Chen}, {\em Pension regulation and the market value of
  pension liabilities: A contingent claims analysis using parisian options},
  Journal of Banking \& Finance, 34 (2010), pp.~1201--1214.

\bibitem{campbell2002strategic}
{\sc J.~Y. Campbell and L.~M. Viceira}, {\em Strategic asset allocation:
  portfolio choice for long-term investors}, Oxford University Press, 2002.

\bibitem{copeland2011target}
{\sc C.~Copeland}, {\em Target-date fund use in 401 (k) plans and the
  persistence of their use, 2007-2009}, EBRI Issue Brief,  (2011).

\bibitem{coxhuang1989}
{\sc J.~C. Cox and C.-F. Huang}, {\em Optimal consumption and portfolio choices
  when asset prices follow a diffusion process}, Journal of Economic Theory, 49
  (1989), pp.~33--83.

\bibitem{crandall1992user}
{\sc M.~G. Crandall, H.~Ishii, and P.-L. Lions}, {\em User's guide to viscosity
  solutions of second order partial differential equations}, Bulletin of the
  American Mathematical Society, 27 (1992), pp.~1--67.

\bibitem{cuoco1997optimal}
{\sc D.~Cuoco}, {\em Optimal consumption and equilibrium prices with portfolio
  constraints and stochastic income}, Journal of Economic Theory, 72 (1997),
  pp.~33--73.

\bibitem{cvitanic2001utility}
{\sc J.~Cvitani{\'c}, W.~Schachermayer, and H.~Wang}, {\em Utility maximization
  in incomplete markets with random endowment}, Finance and Stochastics, 5
  (2001), pp.~259--272.

\bibitem{davies1981uncertain}
{\sc J.~B. Davies}, {\em Uncertain lifetime, consumption, and dissaving in
  retirement}, Journal of Political Economy, 89 (1981), pp.~561--577.

\bibitem{davis2006optimal}
{\sc M.~H. Davis}, {\em Optimal hedging with basis risk}, in From stochastic
  calculus to mathematical finance, Y.~Kabanov, R.~Liptser, and J.~Stoyanov,
  eds., Springer, 2006, pp.~169--187.

\bibitem{DelbaenSchachermayer2006}
{\sc F.~Delbaen and W.~Schachermayer}, {\em The Mathematics of Arbitrage},
  Springer Finance, Springer-Verlag, Berlin, 2006.

\bibitem{duffie1997hedging}
{\sc D.~Duffie, W.~Fleming, H.~M. Soner, and T.~Zariphopoulou}, {\em Hedging in
  incomplete markets with {HARA} utility}, Journal of Economic Dynamics and
  Control, 21 (1997), pp.~753--782.

\bibitem{DybvigLiu2010}
{\sc P.~H. Dybvig and H.~Liu}, {\em Lifetime consumption and investment:
  retirement and constrained borrowing}, Journal of Economic Theory, 145
  (2010), pp.~885--907.

\bibitem{el1998optimization}
{\sc N.~El~Karoui and M.~Jeanblanc-Picqu{\'e}}, {\em Optimization of
  consumption with labor income}, Finance and Stochastics, 2 (1998),
  pp.~409--440.

\bibitem{federico2013utility}
{\sc S.~Federico, P.~Gassiat, and F.~Gozzi}, {\em Utility maximization with
  current utility on the wealth: regularity of solutions to the {HJB}
  equation}, Finance and Stochastics, 19 (2015), pp.~415--448.

\bibitem{federico2012impact}
\leavevmode\vrule height 2pt depth -1.6pt width 23pt, {\em Impact of time
  illiquidity in a mixed market without full observation}, Mathematical
  Finance, 27 (2017), pp.~401--437.

\bibitem{fleming2006controlled}
{\sc W.~H. Fleming and H.~M. Soner}, {\em Controlled Markov Processes and
  Viscosity Solutions}, Stochastic Modelling and Applied Probability, Springer,
  New York, 2006.

\bibitem{hepearson1991a}
{\sc H.~He and N.~D. Pearson}, {\em Consumption and portfolio policies with
  incomplete markets and short-sale constraints: the finite-dimensional case},
  Mathematical Finance, 1 (1991), pp.~1--10.

\bibitem{hepearson1991b}
\leavevmode\vrule height 2pt depth -1.6pt width 23pt, {\em Consumption and
  portfolio policies with incomplete markets and short-sale constraints: The
  infinite dimensional case}, Journal of Economic Theory, 54 (1991),
  pp.~259--304.

\bibitem{horst2014forward}
{\sc U.~Horst, Y.~Hu, P.~Imkeller, A.~R{\'e}veillac, and J.~Zhang}, {\em
  Forward--backward systems for expected utility maximization}, Stochastic
  Processes and their Applications,  (2014).

\bibitem{hu2005utility}
{\sc Y.~Hu, P.~Imkeller, and M.~M{\"u}ller}, {\em Utility maximization in
  incomplete markets}, The Annals of Applied Probability, 15 (2005),
  pp.~1691--1712.

\bibitem{hugonnier2004optimal}
{\sc J.~Hugonnier and D.~Kramkov}, {\em Optimal investment with random
  endowments in incomplete markets}, The Annals of Applied Probability, 14
  (2004), pp.~845--864.

\bibitem{koo1998consumption}
{\sc H.-K. Koo}, {\em Consumption and portfolio selection with labor income: a
  continuous time approach}, Mathematical Finance, 8 (1998), pp.~49--65.

\bibitem{krylov1987nonlinear}
{\sc N.~V. Krylov}, {\em Nonlinear {E}lliptic and {P}arabolic {E}quations of
  the {S}econd {O}rder}, Mathematics and its applications ({S}oviet series),
  Reidel, Dordrecht, 1987.

\bibitem{krylov2009controlled}
\leavevmode\vrule height 2pt depth -1.6pt width 23pt, {\em Controlled
  {D}iffusion {P}rocesses}, vol.~14 of Stochastic Modelling and Applied
  Probability, Springer-Verlag, Berlin, 2009.

\bibitem{merton1969}
{\sc R.~Merton}, {\em Lifetime portfolio seclection under uncertainty: the
  continuous-time case}, Review of Economic Statistics, 51 (1969),
  pp.~247--257.

\bibitem{Merton1971}
\leavevmode\vrule height 2pt depth -1.6pt width 23pt, {\em Optimum consumption
  and portfolio rules in a continuous time model}, Journal of Economic Theory,
  3 (1971), pp.~373--413.

\bibitem{Mostovyi2015}
{\sc O.~Mostovyi}, {\em Optimal investment with intermediate consumption and
  random endowment}, Mathematical Finance, 27 (2017), pp.~96--114.

\bibitem{MostovyiSirbu2020}
{\sc O.~Mostovyi and M.~S{\^{\i}}rbu}, {\em Optimal investment and consumption
  with labor income in incomplete markets}, The Annals of Applied Probability,
  30 (2020), pp.~747--787.

\bibitem{pardoux2014stochastic}
{\sc E.~Pardoux and A.~R\u{a}\c{s}canu}, {\em Stochastic differential
  equations, backward {SDE}s, partial differential equations}, vol.~69 of
  Stochastic Modelling and Applied Probability, Springer, Cham, 2014.

\bibitem{pham2009continuous}
{\sc H.~Pham}, {\em Continuous-time stochastic control and optimization with
  financial applications}, vol.~61 of Stochastic Modelling and Applied
  Probability, Springer, Berlin, 2009.

\bibitem{pliska1986}
{\sc S.~R. Pliska}, {\em A stochastic calculus model of continuous trading:
  optimal portfolios}, Mathematics of Operational Research, 11 (1986),
  pp.~371--382.

\bibitem{SonerVukelja2016}
{\sc M.~Soner and M.~Vukelja}, {\em Utility maximization in an illiquid market
  in continuous time}, Mathathematical Methods of Operations Research, 84
  (2016), pp.~285--321.

\bibitem{sundaresan1997valuation}
{\sc S.~Sundaresan and F.~Zapatero}, {\em Valuation, optimal asset allocation
  and retirement incentives of pension plans}, Review of Financial Studies, 10
  (1997), pp.~631--660.

\bibitem{Touzi2012}
{\sc N.~Touzi}, {\em Deterministic and stochastic control, application to
  finance}, lecture notes, Universit\'e Paris 6, 2012.
\newblock available at: \url{http://www.cmap.polytechnique.fr/~touzi}.

\bibitem{wang2008maximal}
{\sc J.~Wang and P.~A. Forsyth}, {\em Maximal use of central differencing for
  {H}amilton-{J}acobi-{B}ellman {PDE}s in finance}, SIAM Journal on Numerical
  Analysis, 46 (2008), pp.~1580--1601.

\bibitem{zariphopoulou}
{\sc T.~Zariphopoulou}, {\em A solution approach to valuation with unhedgeable
  risks}, Finance and Stochastics, 5 (2001), pp.~61--82.

\bibitem{zaslavski2006turnpike}
{\sc A.~J. Zaslavski}, {\em Turnpike Properties in the Calculus of Variations
  and Optimal Control}, Nonconvex Optimization and Its Applications, Springer,
  Boston, 2006.

\bibitem{Zitkovic2011}
{\sc G.~Zitkovic}, {\em Stability of the utility maximization problem with
  random endowment in incomplete markets}, Mathematical Finance, 21 (2011),
  pp.~313--333.

\end{thebibliography}
\end{document}